\documentclass[a4paper,hyper]{JHEP3}
\usepackage{bm}
\input epsf
\def\tw{\textwidth}
\def\putbox#1#2{\epsfxsize=#1\textwidth\epsfbox{#2}}

\def\Eq#1{Eq.~(\ref{#1})}
\def\be{\begin{equation}}
\def\ee{\end{equation}}
\def\bea{\begin{eqnarray}}
\def\eea{\end{eqnarray}}
\def\ptyp{p_{\rm max}}
\def\barf{\bar{f}}

\def\calC{{\mathcal{C}}}
\def\half{\frac{1}{2}}
\def\nc{N_{\rm c}}
\def\Ca{C_{_{\rm A}}}
\def\da{d_{_{\rm A}}}
\def\p{{\bm p}}
\def\x{{\bm x}}
\def\mD{m_{\rm D}}
\def\wpl{\omega_{\rm pl}}
\def\twotwo{2 {\leftrightarrow} 2}
\def\onetwo{1 {\leftrightarrow} 2}
\def\OO{{\cal O}}
\def\ptw{\tilde{p}}
\def\ftw{\tilde{f}}
\def\fet{f_{E_t}}
\def\fel{f_{E_l}}
\def\avEsq{\langle E^2 \rangle}
\def\avBsq{\langle B^2 \rangle}
\def\gsim{\mbox{~{\raisebox{0.4ex}{$>$}}\hspace{-1.1em}
        {\raisebox{-0.6ex}{$\sim$}}~}}
\def\lsim{\mbox{~{\raisebox{0.4ex}{$<$}}\hspace{-1.1em}
        {\raisebox{-0.6ex}{$\sim$}}~}}
\def\Tr{{\rm Tr}\:}

\title{UV Cascade in Classical Yang-Mills Theory}
\author{
Aleksi Kurkela\footnote{kurkela@physics.mcgill.ca},
Guy D. Moore\footnote{guymoore@physics.mcgill.ca} \\
McGill University Department of Physics,\\
3600 Rue University, Montr\'eal, QC, H3A 2T8
}

\abstract{
We study the real-time behavior of classical Yang-Mills theory under
initial conditions with nonperturbatively large, infrared field
amplitudes.  Our lattice study confirms the cascade of energy towards
higher momenta and lower occupancy, which occurs via a scaling
solution
$f[p,t_1] = (t_0/t_1)^{\frac 47}\, f[p (t_0/t_1)^{\frac 17},t_0]$.
Above a characteristic scale $\ptyp$, $f$ falls exponentially;
below $\ptyp$, $f[p] \propto p^{\frac{-4}{3}}$.  We find no evidence for
different infrared exponents or for infrared occupancies in excess of
those described by this scaling solution.  We also investigate what the
fate of large occupancies would be, both in the electric and the
magnetic sector.
}

\preprint{INT-PUB-12-032}

\keywords{Lattice, Classical Yang-Mills, gauge fixing, screening}

\begin{document}


\section{Introduction}

\label{intro}

Consider classical Yang-Mills theory.  If the initial conditions contain
field energy in some region of space with vacuum around, we know that
after a while the energy will be in an expanding nearly-spherical
shell of outward moving waves.  But if the initial conditions are
statistically spatially homogeneous, the fields will re-interact indefinitely.
Now classical Yang-Mills theory, like all nonlinear classical field
theories, has no equilibrium; there is infinite phase space in the deep
ultraviolet.  So the energy should cascade into the ultraviolet, spreading
with ever smaller amplitude over ever more field modes.  But exactly
how does this cascade actually occur?

Recently there has been renewed interest in this question, because
classical Yang-Mills theory should describe the behavior of weakly-coupled,
quantum Yang-Mills theory when the mean occupancy is high, $f(k)\gg1$ for
the most important $k$ in a system.  The initial conditions after a heavy
ion collision, in the limit of large and high energy nuclei, are expected
to involve high-occupancy, effectively classical fields \cite{MVmodel}.
It is not clear to us how realistic the classical field description is
phenomenologically; also, the initial conditions for a heavy ion collision
are spatially inhomogeneous and anisotropic.  Nevertheless, we will
consider the homogeneous and isotropic case as an interesting warmup
problem and lesson in the behavior of classical Yang-Mills theory.
At minimum we believe that it is interesting to study the behavior
of classical Yang-Mills theory in its own right, especially to establish
its differences from scalar field theory.

Quite recently two papers have considered essentially the problem we
raise; Yang-Mills theory at weak coupling and with initial conditions
with high enough typical occupancy that the behavior is that of classical
fields (at least for some period of time).  Two of us argued \cite{KM1}
that the cascade into the ultraviolet would proceed with the dominant
momentum scale $\ptyp$ (the scale where most of the energy density resides)
growing with time as $\ptyp \propto t^{\frac 17}$.  The
occupancy at this scale would behave as
$f(\ptyp,t) \propto t^{-\frac 47}$, and the occupancy for
$p \ll \ptyp$ would scale as $f(p) \propto p^{-1}$.  (The same scaling
for $\ptyp$ was found for scalar field theory much earlier by Micha and Tkachev
\cite{Micha:2004bv}.)
On the same day, a paper by Blaizot, Gelis, Liao, McLerran and Venugopalan
\cite{BGLMV} came to similar conclusions, except that they argued that
there could in addition be the formation of a condensate of infrared
excitations, carrying most of the particle number, but a minority of the
energy density, in the system.  A recent numerical study of the problem
by Berges, Sexty and Schlichting \cite{BSS}
did not try to determine the scaling of momentum or occupancy with time,
but investigated the infrared tail of the spectrum.  Contrary to the
arguments of \cite{KM1,BGLMV}, they found
$f(p) \propto p^{-\alpha}$, with $\alpha=3/2$ at early times and
transforming to $\alpha=4/3$ at late times (an exponent first proposed
in Reference \cite{Berges:2008mr}).  They found no direct evidence for a
condensate, but argued that the $\alpha=3/2$ scaling might be indicative
of a condensate's physical effects.

We will re-examine this problem, using lattice methods.
We begin by reviewing the arguments for
$\ptyp \propto t^{\frac 17}$ and $f \propto t^{-\frac 47}$ scaling,
laid out in \cite{KM1,BGLMV}.
Then we study the evolution directly on the lattice by initializing
large volume lattice systems with random, infrared-dominated and
nonlinear-amplitude initial conditions.  We study their time evolution
over long times, and with detailed control of lattice and initial
condition effects.  We verify that the
$\ptyp \propto t^{\frac 17}$ and $f\propto t^{\frac{-4}{7}}$ scaling is
established very fast, but the infrared behavior with the $p^{-\frac 43}$
scaling emerges somewhat slower; and the interpretation of the infrared
region is complicated by the physics of screening.  Once the physics of
screening is taken into account, we see no evidence for infrared scaling
exponents other than $-4/3$.  We do not see infrared occupancy in excess
of that predicted by a $p^{\frac{-4}{3}}$ exponent, but we nevertheless
study what would be the properties and fate of electric (plasmon) and of
magnetic condensates in the infrared; in each case we find that they
would be short-lived.

\section{Scaling in Classical Gauge Theory}

\label{sec:boltzmann}

We will start by explaining the relation between classical Yang-Mills
theory and the usual quantum theory at weak coupling, and we will
introduce the scales in the problem.  Then we see how the scaling
behavior naturally arises by considering the structure of the scattering
terms in the Boltzmann equations.  This section reviews known arguments,
see for instance Refs \cite{KM1,BGLMV}; readers familiar with its
contents may want to skip this section, except for \Eq{def_Q} and
\Eq{ptyp_def}, which define our scale $Q$ and momentum $\ptyp$.

\subsection{Classical Yang-Mills theory and scales}

\label{scales}

Without $\hbar$, length and energy scales are distinct, and the gauge
coupling is dimensionful.  The easiest way to think about the classical
theory, for someone familiar with the quantum one, is to assume all
occupancies are of order $f(p) \sim 1/g^2$, so $g^2 f(p)$ is a number
of order unity.  Then one takes $g^2$ small (to zero) holding
$g^2 f(p)$ fixed.  This immediately means we can ignore fermions; since
we assume no scalar matter, we will work with pure-glue QCD.
Occupancies are naturally $f(p) \sim 1/g^2$.  If $Q$ is a scale
characterizing the wave-vectors of initial fluctuations, then the energy
density is naturally $\varepsilon \sim Q^4 / g^2$; it is better
to interpret $Q,p$ as wave numbers or inverse lengths than as energy or
momentum scales.  ($1/g^2$ plays the role, dimensionally, of $\hbar$,
turning an inverse length $Q$ into an energy scale $Q/g^2$.)
When $g^2 f(p)$ is small, the theory behaves nearly
linearly; when $g^2 f(p) \gsim 1$ the behavior is highly nonlinear, and
the notions of wave-vector and occupancy become highly gauge dependent.
Here ``small'' does {\sl not} mean $\propto g^2$, since we take $g^2$
small first.  Rather, the behavior is perturbative when there is some
other expansion parameter $\lambda$, with $f(p) \sim \lambda/g^2$,
and $\lambda \ll 1$ but $\lambda \gg g^2$.  We will see that the inverse
system age -- actually $t^{-\frac 47}$ -- plays the role of $\lambda$.

We will consider extensive systems with a mean energy density
$\varepsilon \sim Q^4/g^2$.  To make quantitative statements easier,
we will {\sl define} the scale $Q$ in terms of $\varepsilon$ as follows.
When Yang-Mills theory is nearly linear, it makes sense to fix the gauge
and describe it in terms of quasiparticle excitations with occupancy
$f(p)$, in terms of which the energy density would be%
\footnote{%
    Here we assume we are considering SU$(\nc)$ gauge theory; for a general
    gauge group, replace $(\nc^2{-}1)$ with the dimension of the
    adjoint representation $\da$ and $\nc$ with the second Casimir of
    the adjoint representation $\Ca$.}
\bea
\varepsilon &\simeq & \sum_{sc}
      \int \frac{d^3 k}{(2\pi)^3} \: k\: f(k)
\nonumber \\
g^2\nc \varepsilon & \simeq &
  \frac{2(\nc^2{-}1)}{2\pi^2} \int k^3 \; \left[g^2 \nc f(k)\right] \: dk \,,
\label{energy}
\eea
where the sum $\sum_{sc}$ is a sum over spin and color states,
$\sum_{sc} 1 = 2(\nc^2{-}1)$.  With this in mind, we simply define the
scale $Q$ as
\be
g^2 \nc \varepsilon =  \frac{2(\nc^2{-}1)}{2\pi^2} \; Q^4  \qquad
\mbox{or} \qquad
Q \equiv \left( \frac{\pi^2 g^2 \nc \, \varepsilon}
                     {\nc^2{-}1} \right)^{\frac 14}
\label{def_Q}
\ee
so that, in the perturbative regime, we have
\be
Q^4 \simeq \int k^3 \;g^2 \nc f(k)\; dk \,.
\label{Qvsf}
\ee
We will generally use dimensionless time $Qt$ and dimensionless wave
numbers $p/Q$.

\subsection{Scaling from the Boltzmann equation}

\label{Scaling}

Now consider the Boltzmann equation describing the time evolution of the
occupancy $f(p)$.  According to Arnold, Moore, and Yaffe \cite{AMY5}, it
is of generic form
\begin{eqnarray}
\frac{\partial f(p,t)}{\partial t} &=& - \calC_{\twotwo}[f(p)]
              -\calC_{\onetwo}[f(p)] \,,
\nonumber \\
\calC_{\twotwo}[f(p)] & = & \frac{1}{2p}
  \int_{k;p'k'} |\overline{\cal M}_{pk;p'k'}|^2 \;
  (2\pi)^4 \delta^4(p{+}k{-}p'{-}k') \times
\nonumber \\ && \hspace{1.4cm}
  \Big( f(p) f(k)[1{+}f(p')][1{+}f(k')] -
        [1{+}f(p)][1{+}f(k)] f(p') f(k') \Big) \quad
\label{Boltzmann1}
\end{eqnarray}
(we discuss $\calC_{\onetwo}$ in a moment).  Here the
phase space integrals are $\int_k = \int \frac{d^3 k}{(2\pi)^3 2k^0}$.
For pure-glue QCD we have
\be
|\overline{\cal M}_{pk;p'k'}|^2 = 4 \nc^2 g^4 \left( 3 - \frac{su}{t^2}
-\frac{st}{u^2} - \frac{tu}{s^2} \right)
\label{Msq}
\ee
with $s,t,u$ the usual Mandelstam variables, and with the
$su/t^2$ and $st/u^2$ matrix elements IR regulated as described in
Ref \cite{AMY5}.  We are interested in the behavior for
$f\sim 1/g^2 \gg 1$, which allows a little simplification.  The terms
in \Eq{Boltzmann1} proportional to $f^4$ cancel between ``loss'' and
``gain'' terms (the first and second terms in the large parenthesis in
the last line); but the terms of order $f^3$ do not; and since $f^3 \gg
f^2$ we can drop the $f^2$ terms and write
\bea
&&  \Big( f(p) f(k)[1{+}f(p')][1{+}f(k')] -
        [1{+}f(p)][1{+}f(k)] f(p') f(k') \Big)
\nonumber \\
& \simeq &
        f(p) f(k) f(p') f(k') \Big( f^{-1}(p') + f^{-1}(k')
                             -f^{-1}(p) - f^{-1}(k) \Big)
\qquad \mbox{for $f\gg 1$}\,.
\label{occupancies}
\eea
For generic values of $f$ and for large momentum transfers%
\footnote{%
    When $\p'\simeq \p$ so $-t\ll s$, ${\cal M}^2$ diverges as $s^2/t^2$
    but there are
    compensating cancellations in \Eq{occupancies}, rendering the
    collision term at worst logarithmically divergent, cut off by
    screening effects.}
so $s\sim -t \sim -u$, there are no cancellations in the round bracket
in the last line, and this term is of order $f^3$.
Next, \Eq{Msq} shows that $|{\overline{\cal M}}|^2$ is of order $g^4$.
Therefore both sides of \Eq{Boltzmann1} are of order $g^{-2}$.
Since $|{\overline{\cal M}}|^2$ is of order $\nc^2 g^4$, it is
natural to assume $f(p) \propto 1/(g^2 \nc)$.  If we introduce
$\barf = g^2 \nc f$, then all factors of $g^2 \nc$ cancel when we
express the Boltzmann equations in terms of $\barf$.
An explicit factor of $g^2 \nc$ also cancels the
factor associated with $f(p)$ in determining the screening mass, needed
in the IR regulation of ${\cal M}^2$.  Since the leading order
expressions for $\calC_{\onetwo}$ and $\calC_{\twotwo}$ do not contain
subleading-in-$1/\nc$ dependence, there is also no dependence on $\nc$
left, which means that a study using the group SU(2) should return the
same scaling behavior as any other SU($\nc$).

With $g^2$ taken care of, we now examine the expected scaling properties
with time.  It is natural to assume that, a time $t$ after some initial
conditions are established, and considering a typical momentum in the
range which dominates the system energy density, the time derivative of
the occupancy is of order%
\footnote{
    If $df/dt \gg f/t$ then the occupancy quickly adjusts to be
    something different; and $df/dt \ll f/t$ can only occur at early
    times for low-occupancy initial conditions.  If we are interested in
    late times, the absence of a thermal ensemble ensures that the
    occupancies will eventually show evolution with $df/dt \sim f/t$.}
\be
\frac{\partial \barf(p,t)}{\partial t} \sim \frac{\barf}{t} \,.
\label{dfdt}
\ee
Since the matrix element is dimensionless, in terms of the (time
dependent) typical momentum scale $\ptyp(t)$, the momentum and
$f$-scaling of the two sides of the Boltzmann equation are
\be
\frac{\barf(\ptyp,t)}{t} \sim \ptyp(t) \barf^3(\ptyp,t) \,.
\label{Cscales}
\ee
This has a simple interpretation; the scattering rate $\Gamma$ for a
typical particle should be $\Gamma t \sim 1$.  The scattering rate
is $\OO( g^4 \ptyp f^2 ) = \OO ( \ptyp \barf^2)$; the factor
$\ptyp$ is on dimensional grounds, $g^4$ is because there are two
vertices involved in a $\twotwo$ scattering process, and
$f^2$ is the occupancy of the scattering target and the stimulation
factor for the final-state scattering target.  (The stimulation factor
for the outgoing particle under consideration cancels between gain and
loss terms.)  The $\calC_{\onetwo}$ term has the same $g^2$ and $f$
scaling.%
\footnote{
    Very briefly, the $\onetwo$ process involves a small-angle
    scattering, with a splitting.  The rate for small-angle scattering
    is $g^4 p f^2$ like $\calC_{\twotwo}$, times a soft enhancement
    $p^2/\mD^2$, which is the ratio of the typical momentum to the
    screening scale $\mD^2 \sim g^2 p^2 f$.  And the splitting
    introduces an additional factor of $g^2 f$, with $f$ the stimulation
    associated with the extra particle.  Putting it together, the rate
    is $g^4 p f^2 (p^2/(g^2 p^2 f))(g^2 f) = g^4 p f^2$, the same as the
    rate of large-angle $\twotwo$ processes.}

Furthermore, energy conservation implies that
\be
g^2 \nc \varepsilon \sim Q^4 = \int p^3 \barf(p)\:dp
  \sim \ptyp^4(t) \barf(\ptyp,t) \quad \Rightarrow \quad
\ptyp^4(t) \barf(\ptyp,t) \sim Q^4
\label{Escales}
\ee
and in particular this combination is time independent.
Solving these two relations, \Eq{Cscales} and \Eq{Escales}, for
$\barf$ and $\ptyp$ in terms of $Q$ and $t$, we find
\bea
\label{pscaling}
\ptyp & \sim & Q (Qt)^{\frac 17} \,,
\\
\label{fscaling}
\barf(\ptyp,t) & \sim & (Qt)^{-\frac 47} \,,
\eea
the scaling behaviors already found in Refs \cite{KM1,BGLMV}
($\Lambda$ of Ref.~\cite{BGLMV} refers to the same parametric scale as
$\ptyp$).

Since we will use $\ptyp$ repeatedly in the following sections, it
behooves us to define it more precisely.  We will define it as
\be
\ptyp^2 \equiv \frac{\langle \Tr ({\bm D}\times {\bm B})^2 \rangle}
                    {\half \langle \Tr ( {\bm B}^2 + {\bm E}^2) \rangle}
\simeq              \frac{\int k^5\: \barf(k) \: dk}
                    {\int k^3 \: \barf(k) \: dk} \,,
\label{ptyp_def}
\ee
where the angular brackets indicate volume (and/or ensemble) averaging.
The definition here is convenient because it is gauge invariant and easy
to evaluate on the lattice; the second expression is the small-amplitude
or quasiparticle behavior of the first expression.

The arguments above suggest the same scaling for scalar fields should
also occur \cite{Micha:2004bv}.  But for
scalars there is an extra complication.  The total particle
number density is $n \sim \int f(p) p^2 dp$, which according to
\Eq{pscaling} and \Eq{fscaling} should scale as
$n\propto t^{\frac{-1}{7}}$.  But $\calC_{\twotwo}$ does not change
particle number, so no scaling solution can actually solve
\Eq{Boltzmann1} if only $\calC_{\twotwo}$ is present.
In a scalar theory this means that the extra particle number
must somehow crowd into the infrared, where it either forms a condensate
or is destroyed by nonlinear processes, which occur faster at the very
high occupancies which will occur in the infrared (for a recent study
see Ref.~\cite{Sexty}).  For us, $\calC_{\onetwo}$ is of the same order
as $\calC_{\twotwo}$, so particle number is not conserved and there need
not be a condensate or enhanced infrared occupancy.  Whether or not such
infrared enhancements occur in practice requires a more detailed
solution of the problem, which we turn to next.

\section{Lattice treatment}

\label{sec:latt}

We will solve the Boltzmann equations for this system in a future
publication.  Here we will directly solve classical Yang-Mills theory
fully nonperturbatively, on the lattice.  Our treatment will emphasize
finding the scaling solution and identifying (transient) corrections to
scaling.  We will also make an effort to investigate the possibility of
infrared condensates.  There are large algorithmic advantages to
considering only SU(2) gauge theory, so this study will strictly work
within SU(2).

The technology for studying real-time, classical fields on the lattice
is decades old; see for instance \cite{Ambjorn}.
The main issues are, what initial conditions should be
studied and how much do results depend on the choice, what physical
measurables should be used, and how does one monitor finite volume and
finite lattice spacing effects?

\FIGURE{
\putbox{0.45}{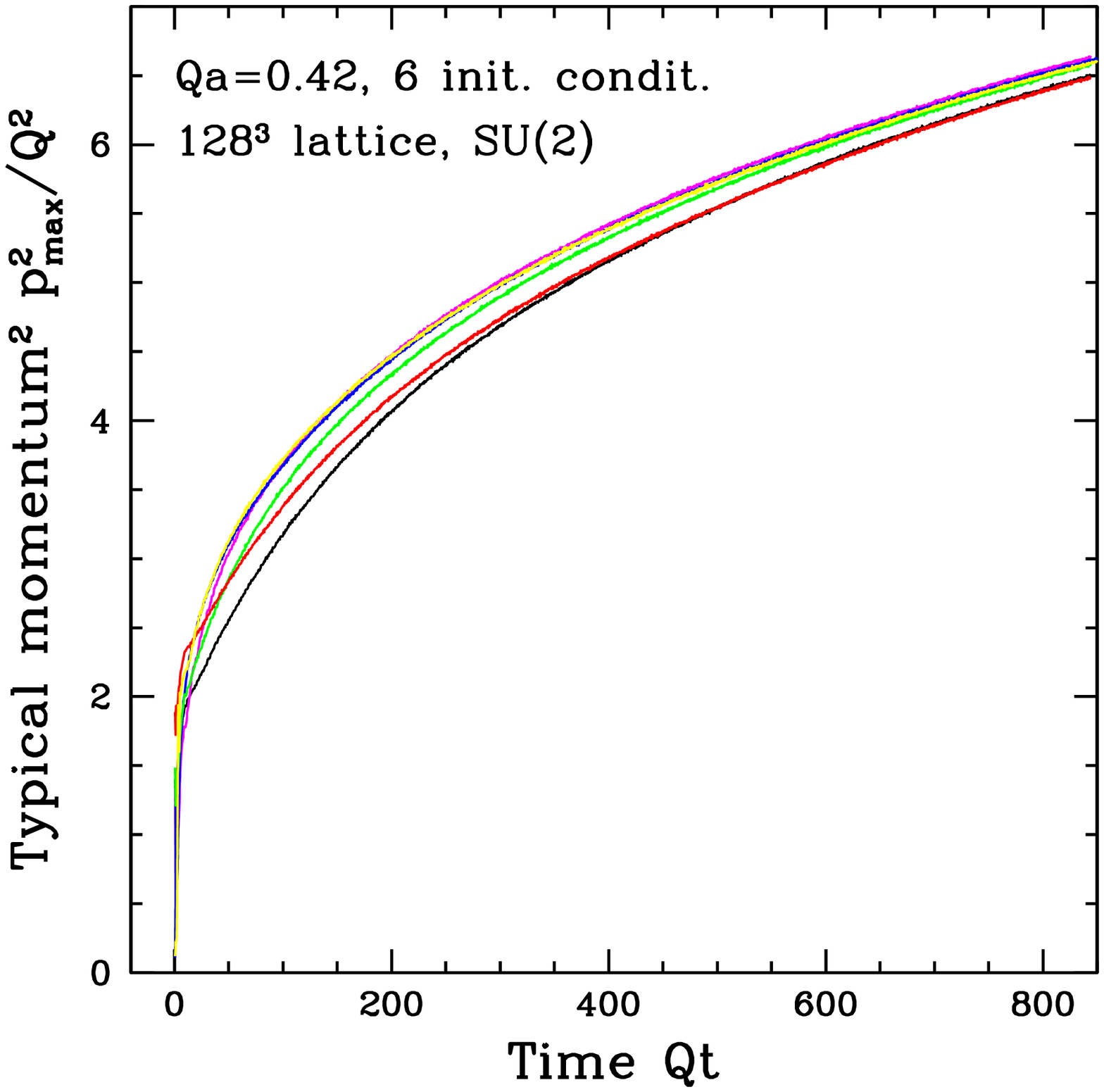}
\hspace{0.05\tw}
\putbox{0.45}{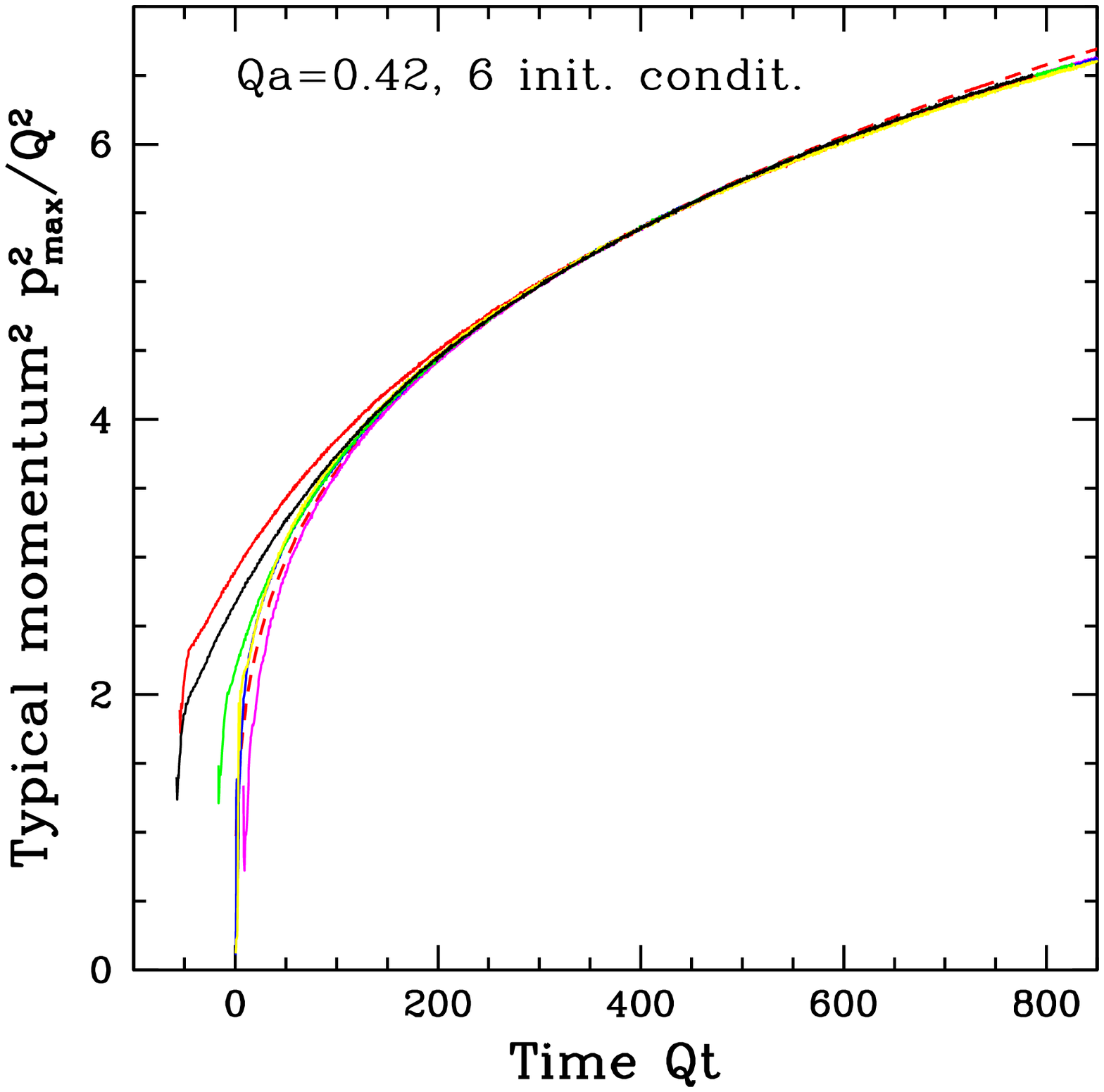}
\caption{
  Evolution of the characteristic momentum scale $\ptyp^2$ as a function
  of time $t$, each scaled by $Q$ to be dimensionless.  At the left,
  time is based on the starting time of each simulation; at the right,
  the starting times are shifted so the late-time behavior is in better
  correspondence.  In the righthand figure, the red dotted line is a fit
  based on strict $\ptyp^2 \propto (Qt)^{\frac 27}$ scaling behavior.}
  \label{fig_ptyp}
}

The simplest gauge invariant measurable is $\ptyp^2$ defined in
\Eq{ptyp_def}.  So to examine the approach to scaling we will first look
at results for this quantity.  To do so we created several different
initial conditions, with either infrared electric fields, magnetic
fields, or both present, and with different ranges of momenta occupied.
To keep lattice systematics common, they all have approximately the same
value of $Qa$, namely $Qa\simeq 0.42$.  The evolution behavior of
$\ptyp^2/Q^2$ as a function of $Qt$ is shown for 6 initial conditions in
Figure \ref{fig_ptyp}.  Each curve shows an initial transient, whose
length depends on the initial condition; but after some time they all
approach a common scaling behavior.  In the lefthand figure the time is
based on the starting time of the simulation; but depending on the
initial condition, there may be a delay before the dynamics starts to
track towards the scaling solution (or it may reach the scaling solution
early, if the initial condition is already quite similar).  Therefore,
to understand whether the evolution really approaches a scaling
solution, it makes more sense to shift each initial time so that the
curves are more similar at late time.  We have done so in the righthand
figure, which shows that the different initial conditions very
accurately fall onto the same scaling behavior.  The red dotted line is
a 1-parameter fit, $\ptyp(t) = c (Qt)^{\frac 27}$, which shows that the
curves are obeying the expected time scaling.  Unfortunately, by
$Qt = 450$ the scale $\ptyp$ has already reached $\ptyp a = 1$,
which means that lattice spacing errors start to occur; at
$\ptyp a=1$ we expect of order $10\%$ errors due to higher-dimension
operators in the lattice action.  So precision quantitative results
would demand a finer lattice (smaller $Qa$).

\FIGURE{
\putbox{0.6}{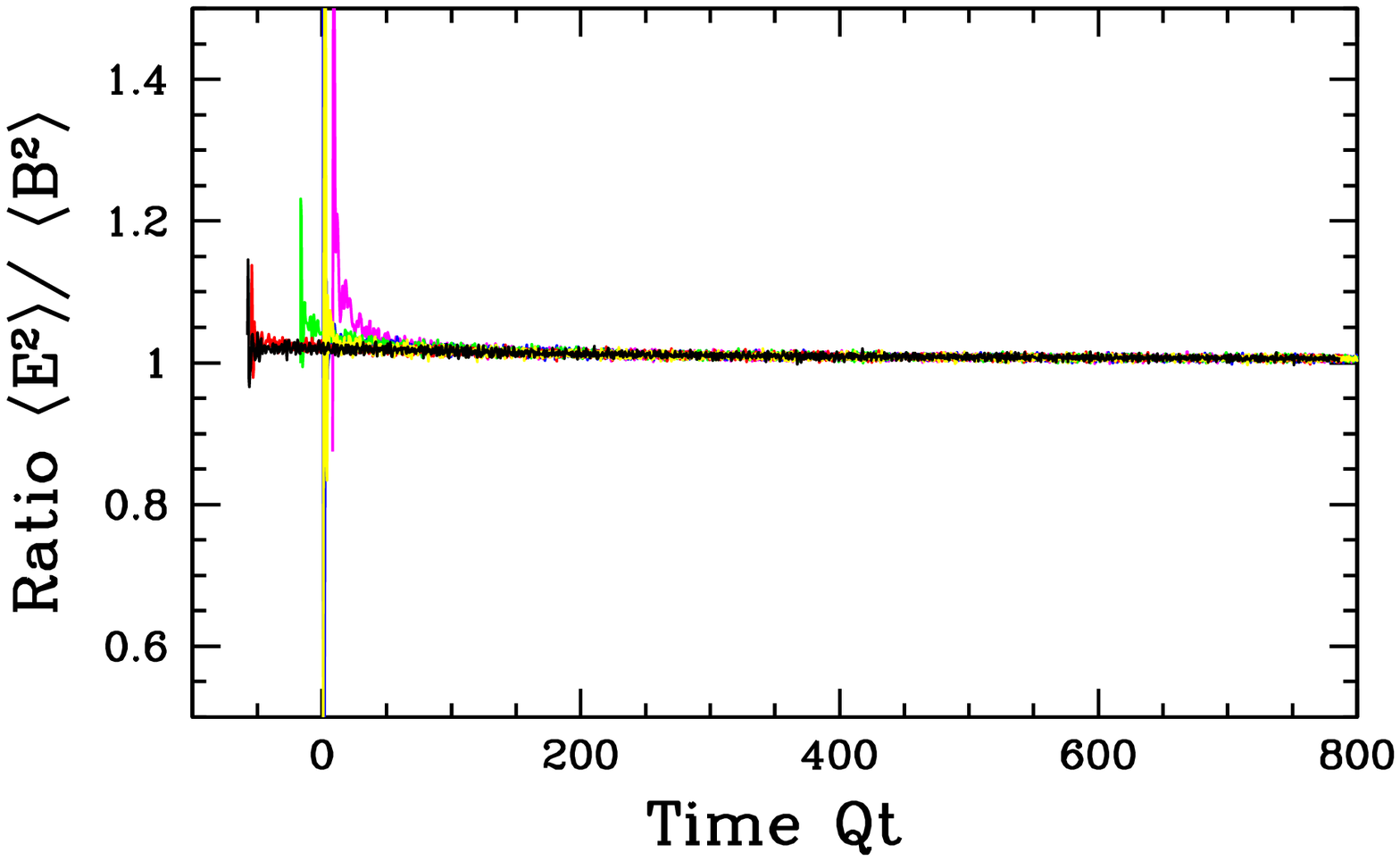}
\caption{
  Ratio of electric to magnetic field energy for the same evolutions as
  in Figure \ref{fig_ptyp}.  Initial transients rapidly decay onto a
  tracking solution with slightly more electric than magnetic energy.}
  \label{fig_eb}
}

We can also look at the gauge invariant ratio of electric to magnetic
energy densities
$\langle \Tr {\bm E}^2 \rangle / \langle \Tr {\bm B}^2 \rangle$, shown
in Figure \ref{fig_eb}.  A quasiparticle with momentum large compared to
the plasma frequency $p \gg \wpl$ should have equal
time-averaged electric and magnetic energy.  Plasmons with
$p < \wpl$ would display more electric than magnetic energy,
whereas magnetic fields in this regime (the Landau cut) would have more
magnetic energy.  So a condensate of plasmon quasiparticles would be
expected to manifest as an excess of electric field energy leading to a
ratio larger than 1, while a condensate of magnetic field would give a
ratio smaller than 1.  In fact, after a short and initial-condition
dependent transient, the ratio settles down to be close to but slightly
above 1, decaying towards 1 with time.%
\footnote{Specifically, in the figure, the blue curve has initial
  conditions with purely electric energy; the yellow curve has purely
  magnetic energy.  The other curves have a mix of electric and magnetic
  but in different momentum ranges.}
This suggests against the existence of large condensates; but we will
examine the evidence in more detail in the next section.

\subsection{Gauge fixed observables}

To learn more about the evolving configurations on the lattice, it is
useful to examine gauge-fixed observables.  Coulomb gauge is the
choice of gauge which minimizes $\int_x {\bm A}^2(x)$.  It is therefore
perhaps the most sensible gauge to use for studying the distribution of
excitations at one moment in time.  And the technology for fixing to
Coulomb gauge on the lattice is very well known, amounting to the fixing
of Landau gauge for the 3-D configuration at an instant
\cite{Mandula}.  Since the gauge fixing procedure emphasizes minimizing
$\int_x {\bm A}^2(x)$ at every time at the expense of minimizing
$A^0$, {\sl unequal} time correlators in this gauge tend to have very
short autocorrelation.  So we will not attempt to untangle unequal time
correlations using gauge fixed methods.

Conventionally, one uses the perturbative relation between occupancy and
field amplitude, for transverse quasiparticle excitations,
\bea
\label{f_and_A}
\int d^3 x \: e^{i\p \cdot \x}
  \langle A_a^i(x) A_b^j(0) \rangle
  &=& \frac{\delta_{ab} {\cal P}_{\rm T}^{ij}(\p)}{|\p|}
       f(p) \,,
\\
\label{f_and_E}
\int d^3 x \: e^{i\p \cdot \x}
  \langle E_a^i(x) E_b^j(0) \rangle
  &=& \Big( \delta_{ab} {\cal P}_{\rm T}^{ij}(\p)\: |\p| \Big)
       f(p)
\eea
(with ${\cal P}_{\rm T}^{ij}=\delta^{ij} - \hat{p}^i \hat{p}^j$ the
transverse projector), to estimate the occupancies as
\bea
\label{A_and_f}
f_{A}(\p) & = & \frac{\delta_{ij} \delta_{ab}}{2(\nc^2{-}1)}
 |\p| \int d^3 x \: e^{i \p \cdot \x}
            \langle A_a^i(x) A_b^j(0) \rangle_{\rm coul}
  \, , \\
\label{E_and_f}
f_{E}(\p) & = & \frac{\delta_{ij} \delta_{ab}}{2(\nc^2{-}1) |\p|}
 \int d^3 x \: e^{i \p \cdot \x}
          \langle E_a^i(x) E_b^j(0) \rangle_{\rm coul} \,.
\eea
In this way we obtain two estimates of the occupancies, one from gauge
fields and one from electric fields.  To illustrate the utility, and the
danger, of this approach, we use it to examine the occupancies of a
nonperturbative lattice system in equilibrium.%
\footnote{%
    The lattice system has an equilibrium because the available momenta
    are restricted to a Brillouin zone, so the phase space is finite.
    From the point of view of continuum classical field theory, this
    equilibrium state is a complete lattice artifact.}
In equilibrium one
usually expects $f(p) = T/p$, which is small provided
$p \ll g^2 \nc T$. To ensure that the occupancy is small for typical
lattice momenta $p \sim \pi/a$, we consider a lattice with spacing
$a = 0.4/(g^2 \nc T)$ (which is $\beta=20$ in the conventional lattice
notation).  The results for the occupancies $f_{A}$ and $f_{E}$ are
shown in Figure \ref{fig_therm}.

\FIGURE{
\putbox{0.8}{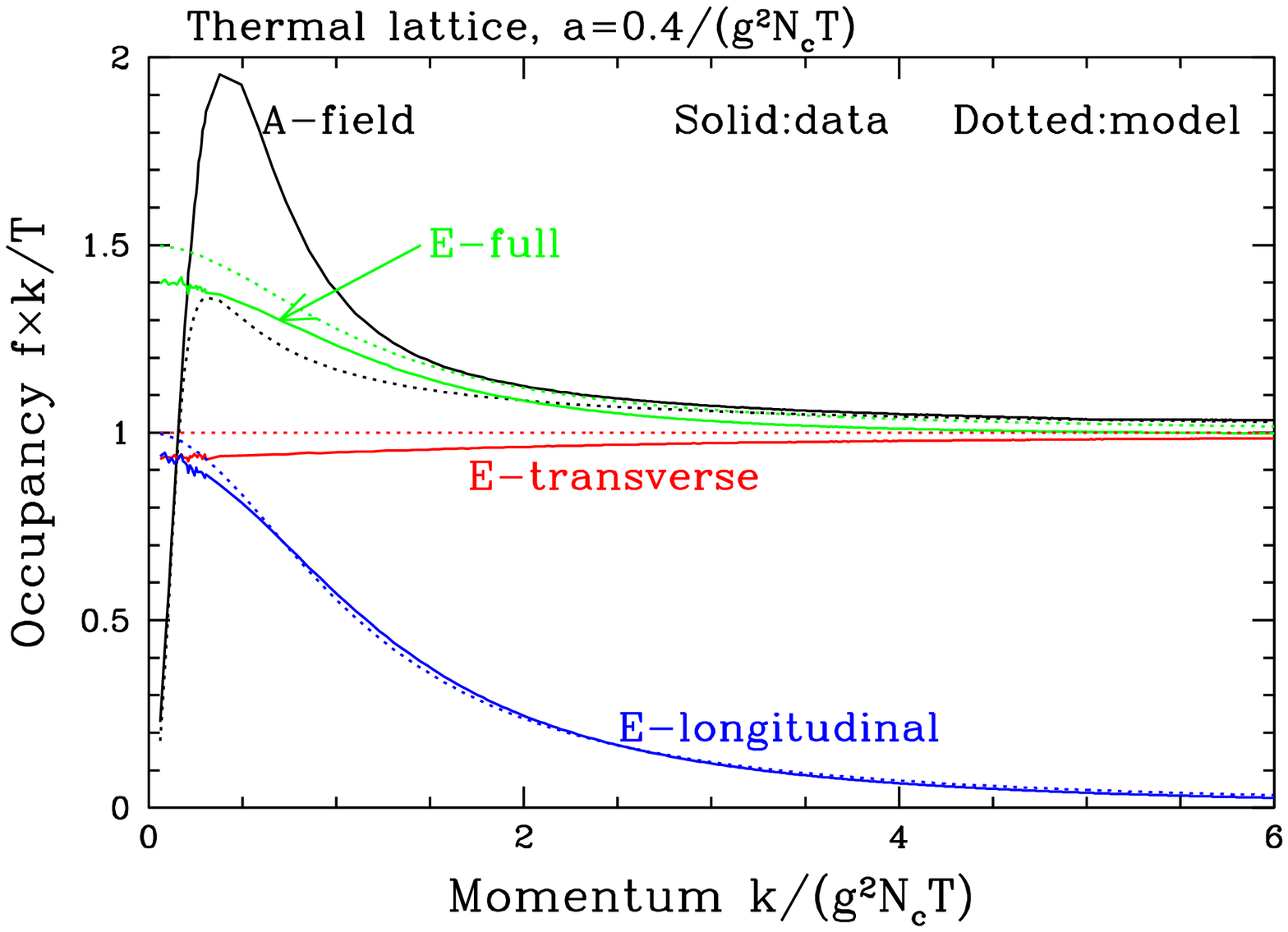}
\caption{
  Equilibrium occupancies in a $256^3$ SU(2) lattice with
  $a=0.4/(g2 \nc T)$, scaled by $k/T$ so at lowest order they should be
  1.  The black curve is $f_{A}$ estimated using the $A$-field
  correlators; green is $f_{E}$ estimated using $E$-field correlators,
  while red and blue are the estimates using the transverse and
  longitudinal components of the $E$-field.  The most faithful estimator
  is the transverse E-field component.}
  \label{fig_therm}
}

The figure shows $f_{A}$ and $f_{E}$ as the black and green curves.  But
whereas the $A$-field is automatically transverse because of our gauge
condition, the electric field is not.  Gauss's law states
${\bm D}\cdot {\bm E}=0$, not $\nabla \cdot {\bm E}=0$, and below the
Debye scale the two become strongly inequivalent.  Therefore we have
separately plotted the transverse electric estimator
\be
\label{Et_and_f}
\fet(\p) = \frac{{\cal P}^{ij}_{\rm T} \delta_{ab}}{2(\nc^2{-}1) |\p|}
 \int d^3 x \: e^{i \p \cdot \x}
            \langle E_i^a(x) E_j^b(0) \rangle_{\rm coul}
\ee
and the longitudinal electric estimator
\be
\label{El_and_f}
\fel(\p) = \frac{\hat{p}^i \hat{p}^j \delta_{ab}}{(\nc^2{-}1) |\p|}
 \int d^3 x \: e^{i \p \cdot \x}
           \langle E_i^a(x) E_j^b(0) \rangle_{\rm coul} \,.
\ee
At the level of hard thermal loops these have simple expressions
(see Appendix \ref{App1})
\be
\label{sumrule}
\fel(\p) = \frac{\mD^2}{\mD^2 + p^2} \frac{T}{p} \,,
\qquad
\fet(\p) = \frac{T}{p} \,.
\ee
The estimator $f_E$ is $f_E = \fet + \half \fel$, which therefore
systematically rises at small momenta $p \lsim \mD$.  Physically, the
reason $f_E$ rises is that, for $p<\mD$, there are three polarization
states contributing to the electric field correlator, but the estimator
\Eq{E_and_f} {\sl assumes} that there are only two.  Therefore, below
the scale $\mD$, the estimator $f_E$ is incorrectly normalized and
systematically high.  However the estimator $\fet$ is based on a
correct counting of degrees of freedom and should be used instead.

The figure also shows that the $A$-field correlator does not reproduce
the expected behavior $f_A(k) = T/k$.  As one goes down in momentum
scale $k$, it first rises above the leading-order estimate, peaks, and
falls to zero at small momentum.  The rise is due to perturbative
corrections to the $AA$ correlator, which are known; at next-to-leading
order,
\be
\int d^3 x \: e^{i\p \cdot \x}
  \langle A_a^i(x) A_b^j(0) \rangle
  = \delta_{ab} {\cal P}_{\rm T}^{ij}(\p)
  \; \frac{1}{p^2 - \frac{11 g^2 \nc T}{64} |\p| + \OO(g^4 T^2)} \,.
\ee
Note that the first correction is suppressed by $1/p$, not by $1/p^2$.
The dotted ``model'' curve shown in Figure \ref{fig_therm} uses the
above correlator with $\OO(g^4 T^2)$ replaced with $g^4 \nc^2 T^2/36$,
which seems to match the infrared behavior.  The ``model'' curves for the
$f_E$ curves are based on \Eq{sumrule} and the lowest-order lattice
value for $\mD^2$, \cite{Kajantie}
\be
\mD^2 a^2 = \frac{2 \nc \Sigma}{\pi \beta}
= \frac{\nc^2 g^2 aT \Sigma}{2\pi} \,, \quad \Sigma=3.1759\ldots \,.
\ee

Because $f_{A}$ rises above the leading-order expectation, using
$\int k^3 f_A(k) dk$ to estimate $\langle \Tr {\bm B}^2 \rangle$
provides a systematic over-estimate; for the lattice we considered, it
over-estimates the actual magnetic energy by 3.5\%.  But $f_E$ gets the
electric energy exactly right.  With these remarks in mind, we will use
$\fet$ to estimate the occupancy and will use $f_A$ as a secondary
estimator, with some caution about the meaning of its infrared
behavior.

\subsection{Approach to scaling}

We have performed a number of classical gauge field evolutions on
lattices of size up to $256^3$ and for times in excess of $Qt=10^4$.
Provided that we concentrate on that part of an evolution where
$a \ptyp \leq 1$, we consistently see the approach to scaling expected
from our kinetic theory study.  To illustrate this, Figure
\ref{fig_scale} shows $g^2 \nc\: f$ versus $p/Q$ at a number of times,
first in absolute units and then building in the assumption of
scaling as described in \Eq{pscaling} and \Eq{fscaling}.  We see that,
after rescaling $f$ and $p$ to remove their dominant time scaling, the
occupancy takes the same form quite accurately over a wide range of
times.

\vspace{1.0em}

\FIGURE{
\putbox{0.45}{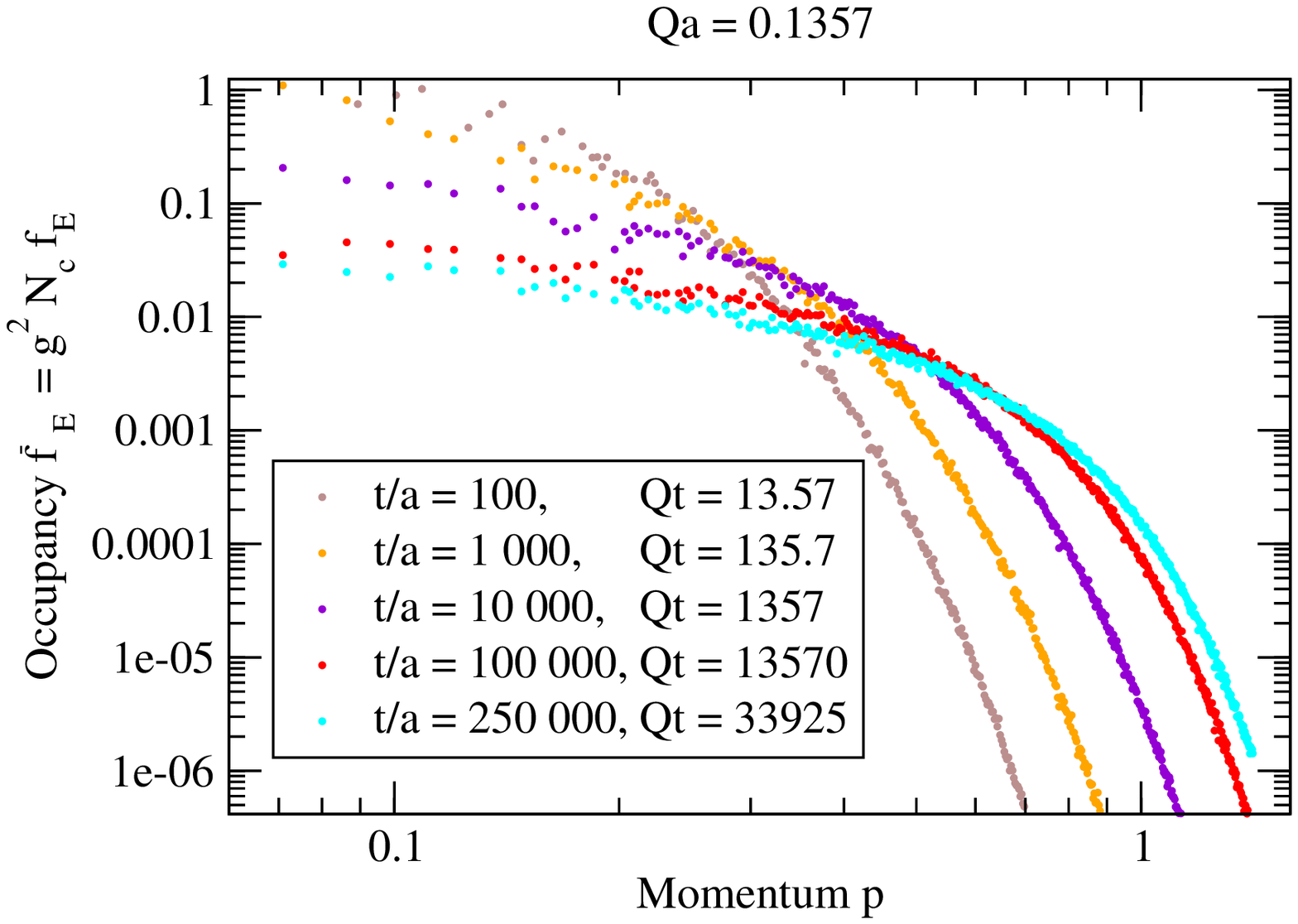}
\hspace{0.05\tw}
\putbox{0.45}{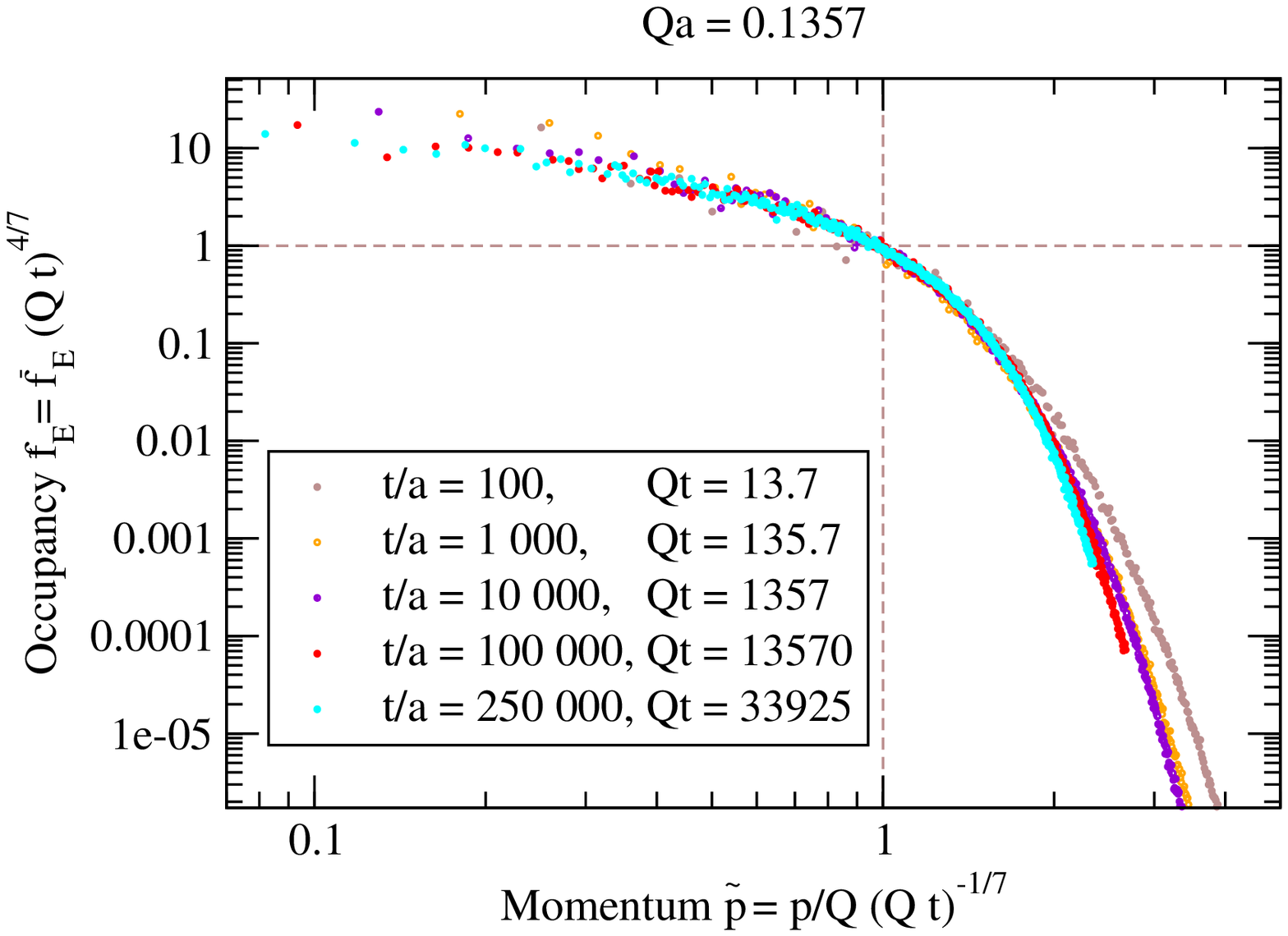}
\caption{
  Left:  occupancy against momentum at a number of times.  Right:  the
  same curves, but scaling out the dominant time dependence of
  $p/Q \propto (Qt)^{\frac 17}$ and
  $g^2 \nc \, f \propto (Qt)^{\frac{-4}{7}}$.}
  \label{fig_scale}
}

How quickly do the gauge fields approach this scaling solution?  To
study this, we look at the same initial conditions used in Figure
\ref{fig_ptyp}.  Figure \ref{fig:approach} shows the occupancies $f_A$
and $\fet$, starting at an early time where the very different
initial conditions are evident and continuing until they have approached
a common behavior.  The spectra agree at the same time that the values
of $\ptyp$ converge in Figure \ref{fig_ptyp}, of order $Qt=60$.

\FIGURE{
\putbox{0.3}{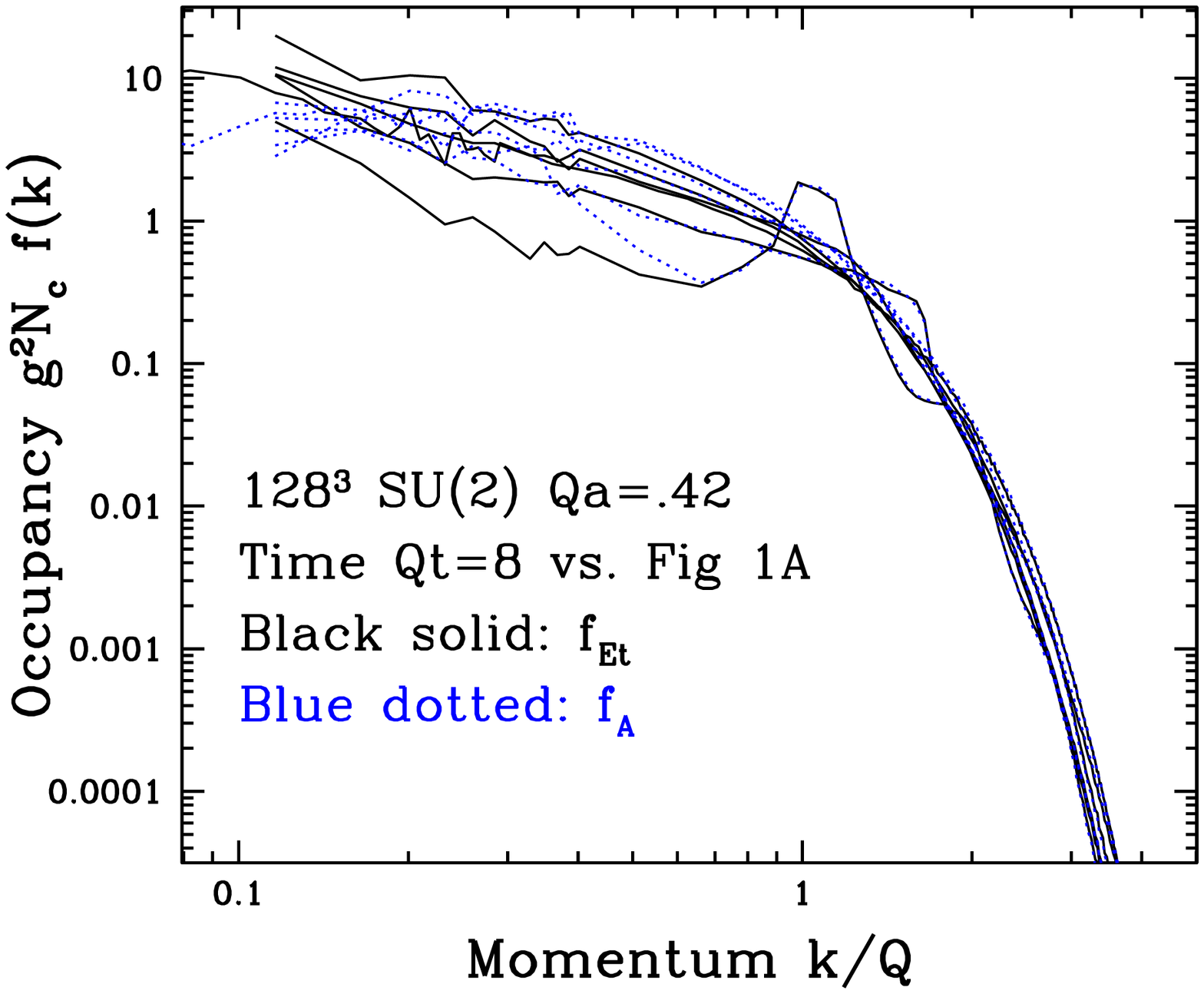}
\hspace{0.02\tw}
\putbox{0.3}{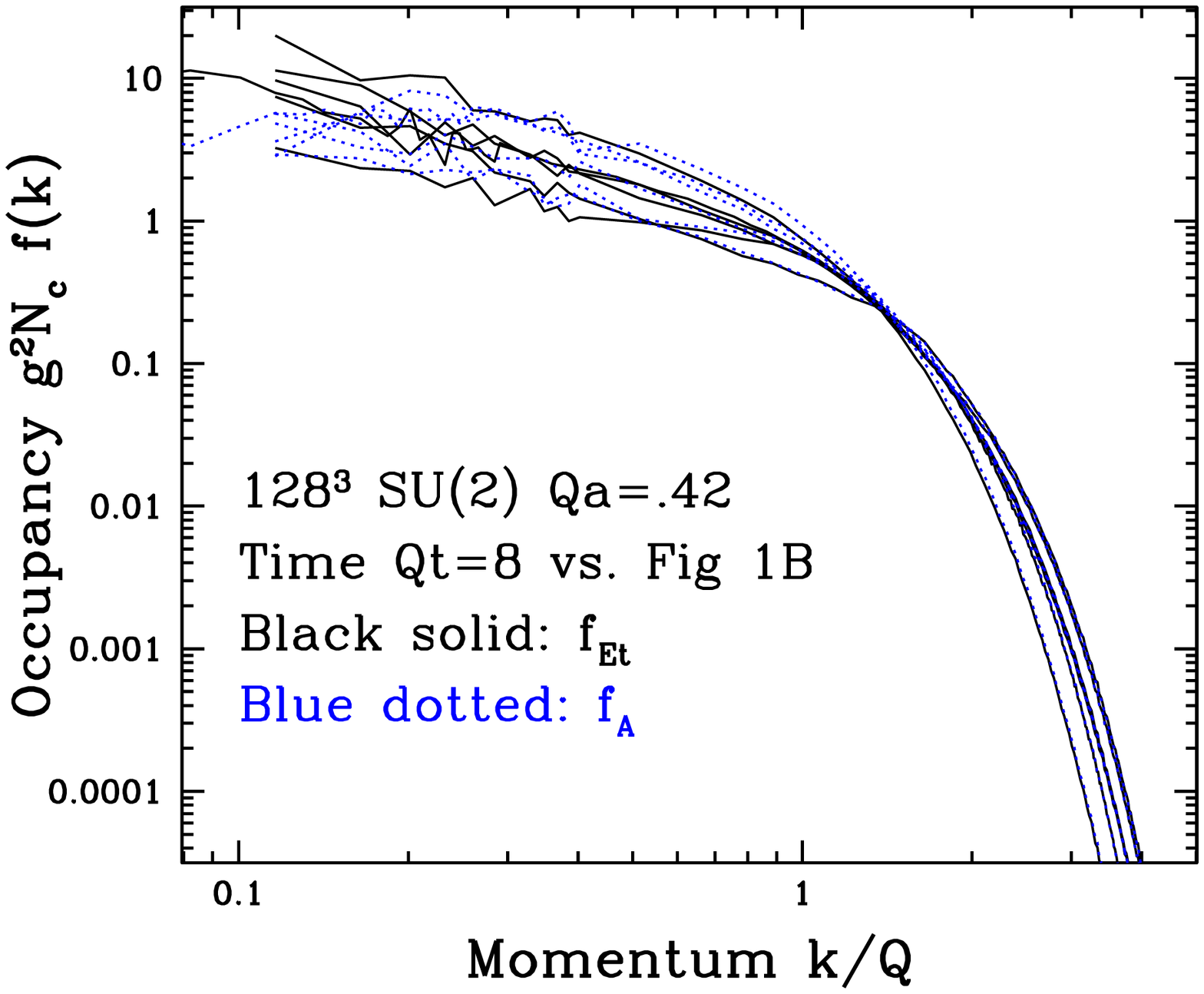}
\hspace{0.02\tw}
\putbox{0.3}{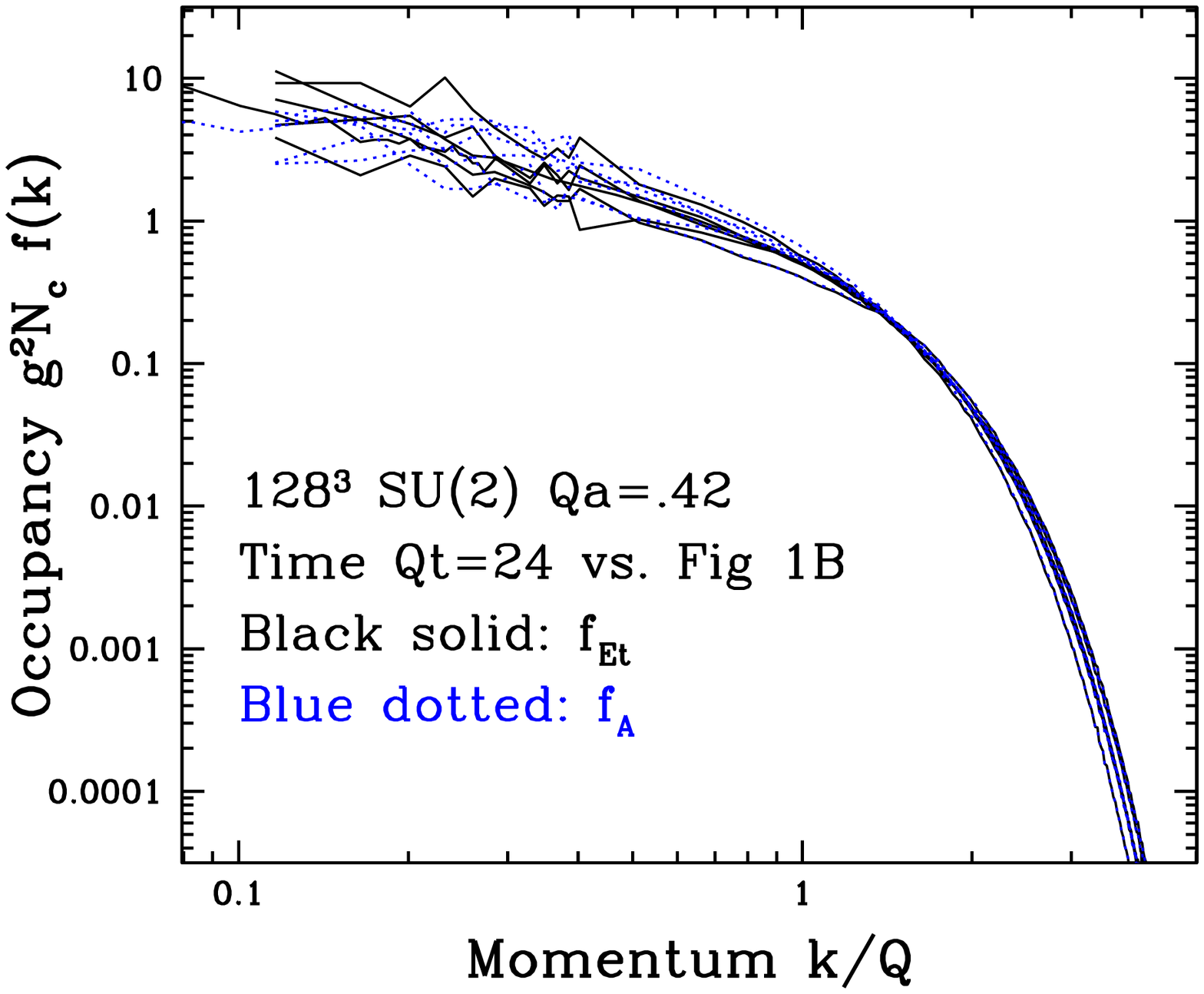} \\
\putbox{0.3}{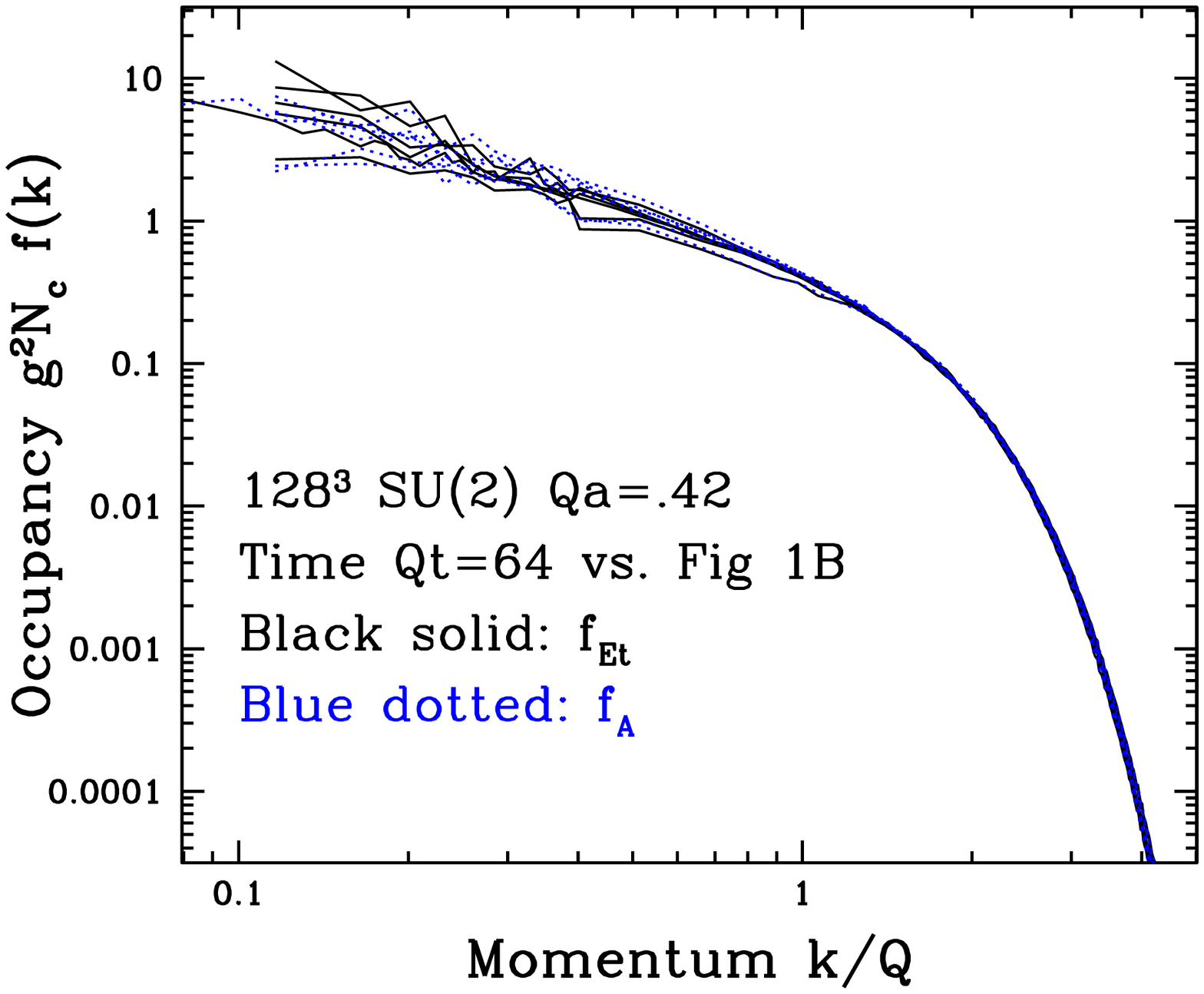}
\hspace{0.02\tw}
\putbox{0.3}{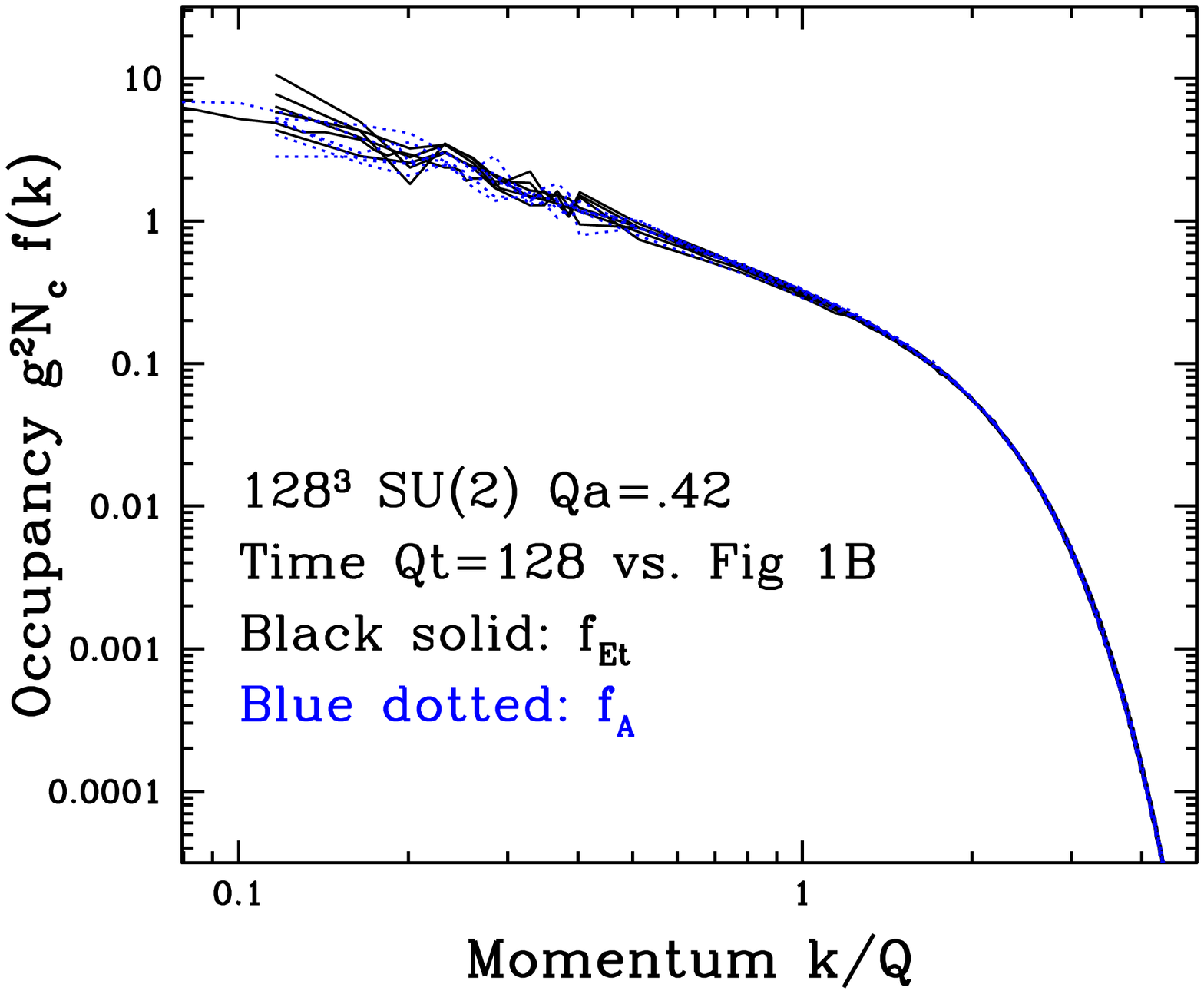}
\hspace{0.02\tw}
\putbox{0.3}{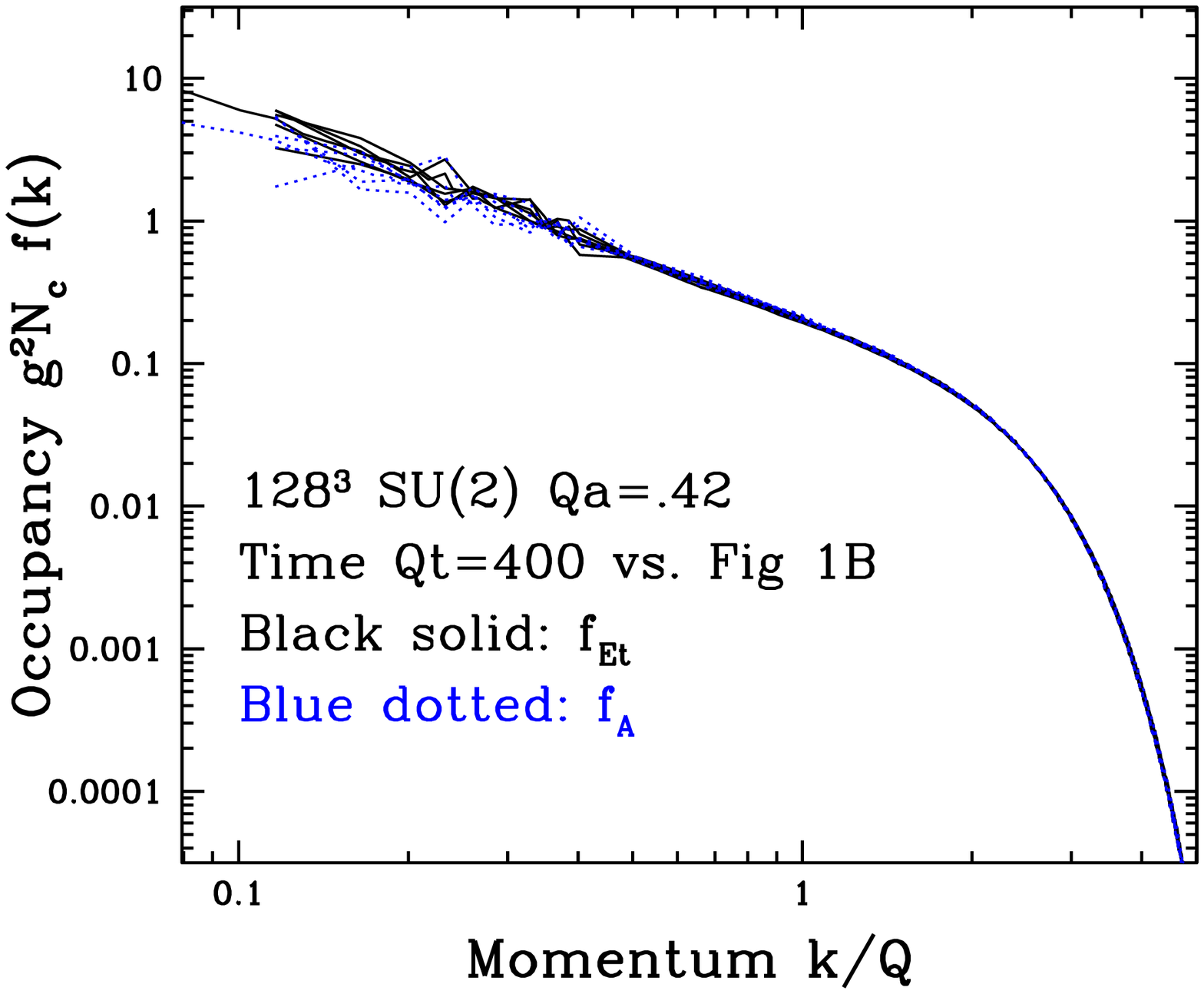}
\caption{\label{fig:approach}
  Approach to the scaling solution for the same 6 initial conditions
  considered in Figure \ref{fig_ptyp}.  Initial differences are notable
  at $Qt=8$ but already small at $Qt=24$ and gone by $Qt=64$.
  Differences in the infrared in the last line are of order statistical
  error.}
}

Looking at Figure \ref{fig_scale} and Figure \ref{fig:approach}, it
appears that the scaling solution is not completely time independent,
but shows some weak time evolution.  First, the large-momentum,
low-occupancy falloff appears to evolve with time.  Second, the infrared
behavior, below the characteristic scale $\ptyp$, also shows some
evolution.  The evolution at low momentum is real and we will look at it
in more detail in the next subsection.  But the evolution at large
momentum is a lattice artifact, caused by higher dimension operators
which become important at momentum scales $p \gsim 1/a$.  To illustrate
this, we plot the UV falloff at several times $Qt$ and lattice spacings
$a \ptyp$ in Figure
\ref{fig:tail}.  The figure shows clearly that the differences between
curves are associated with their taking different values of
$a \ptyp$, not with their being at different times $Qt$.%
\footnote{%
    We should clarify that the momentum in the figure is really the
    square root of the free lattice dispersion $\sqrt{\tilde k^2}$,
    which for our Wilson lattice action is
    $\tilde k^2 = \frac{4}{a^2} \sum_i
              \sin^2 \left( \frac{k_i a}{2} \right)$.}

\FIGURE{
\putbox{0.7}{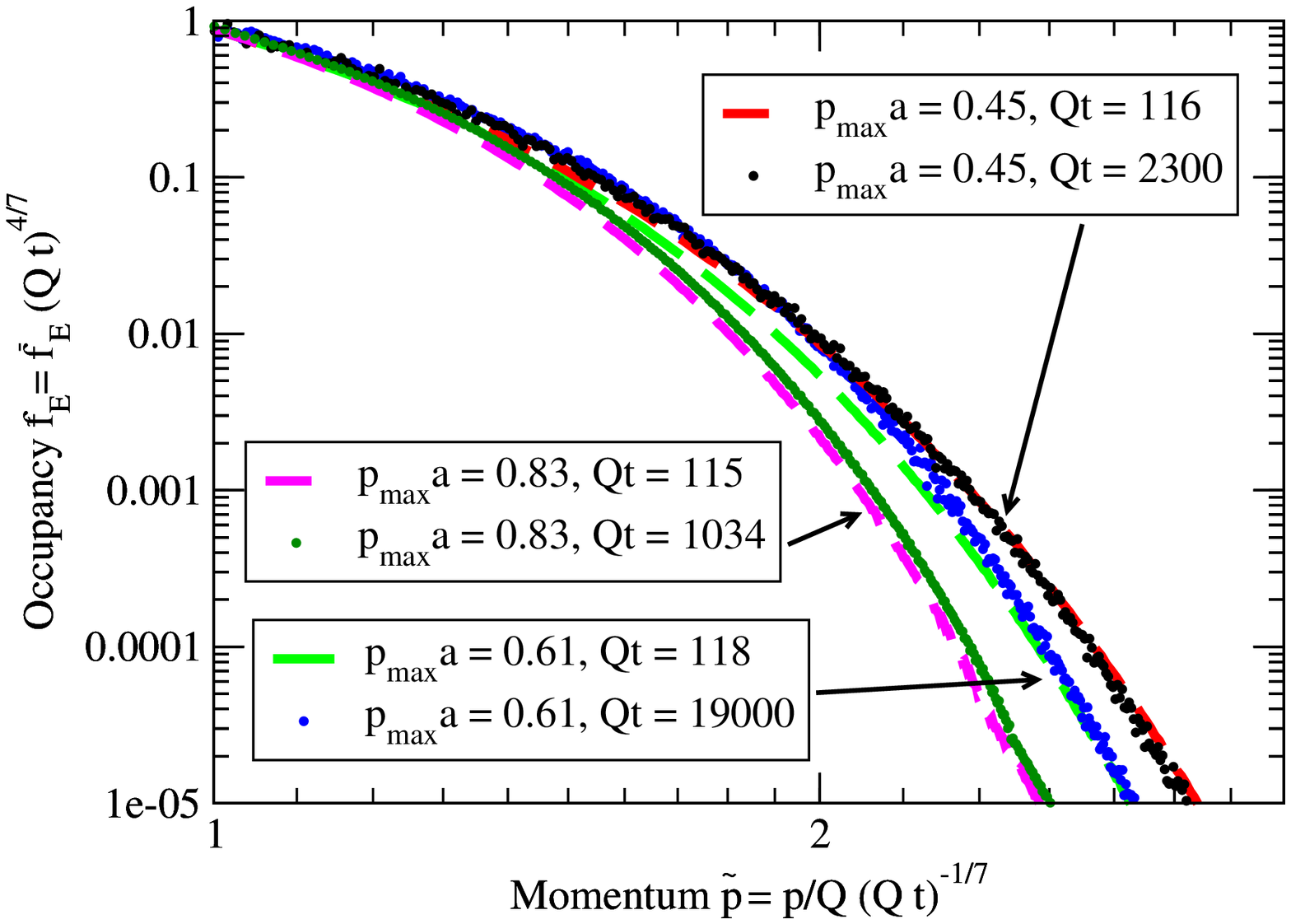}
\caption{
  Ultraviolet falloff from several evolutions, at different times and
  different lattice spacings.  The UV behavior is mostly sensitive to
  $a \ptyp$; at equal $a\ptyp$ it is almost $Qt$ independent.}
\label{fig:tail}
}

\subsection{Evolution of the scaling solution}

Now we examine in more detail whether there is time evolution in the
infrared part of the scaling solution.  We saw in Figure
\ref{fig:approach} that distinct initial conditions approach a common
scaling solution by $Qt=60$.  Our focus will be on time scales this
long or longer.  To do so we performed several independent evolutions on
large ($256^3$) and relatively coarse ($Qa=0.422$) lattices, averaging
and obtaining ensemble error bars.  For times longer than $Qt=500$ we
also performed simulations at $Qa = 0.298$.

\FIGURE{
\putbox{0.45}{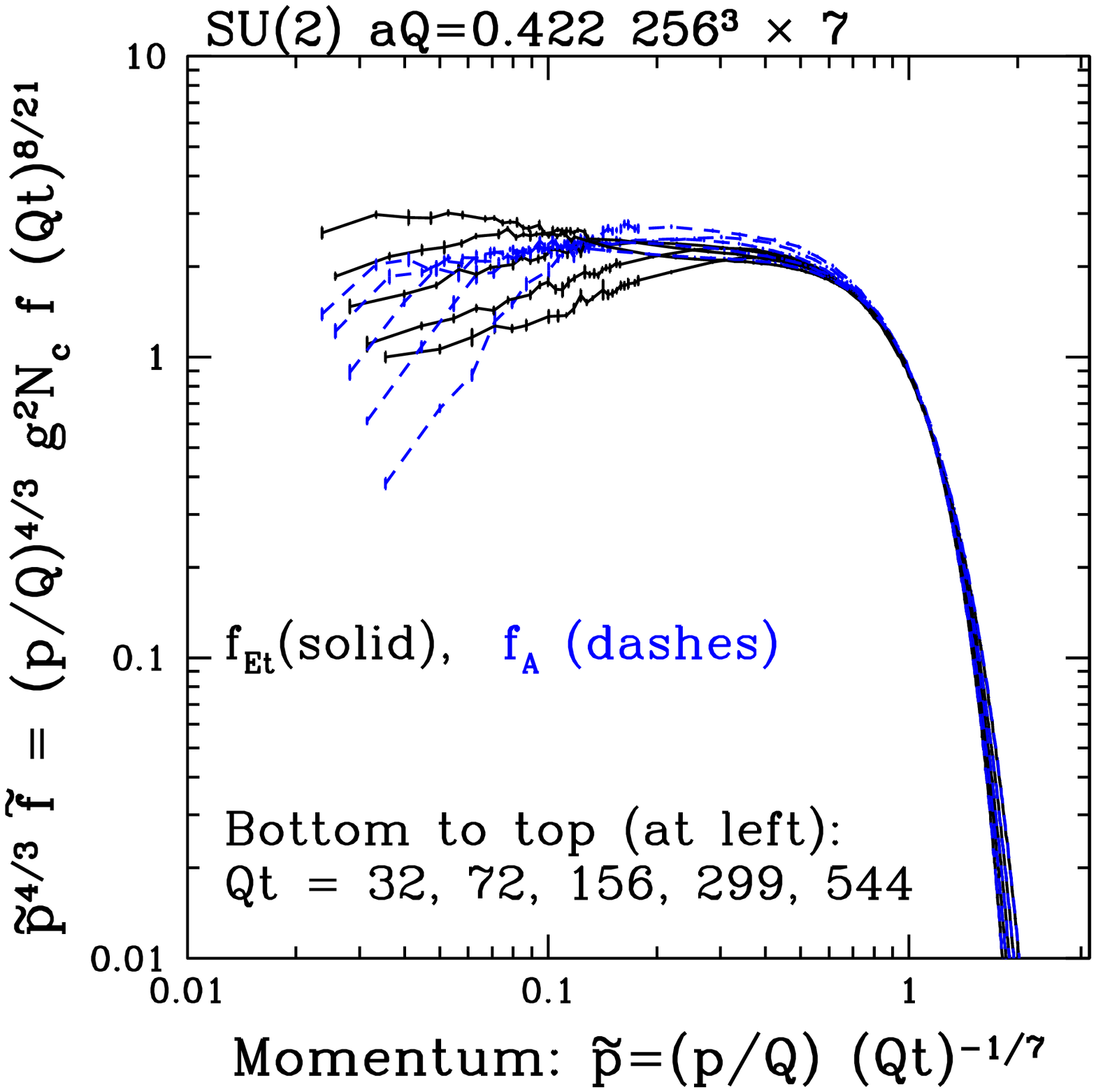} \hfill
\putbox{0.45}{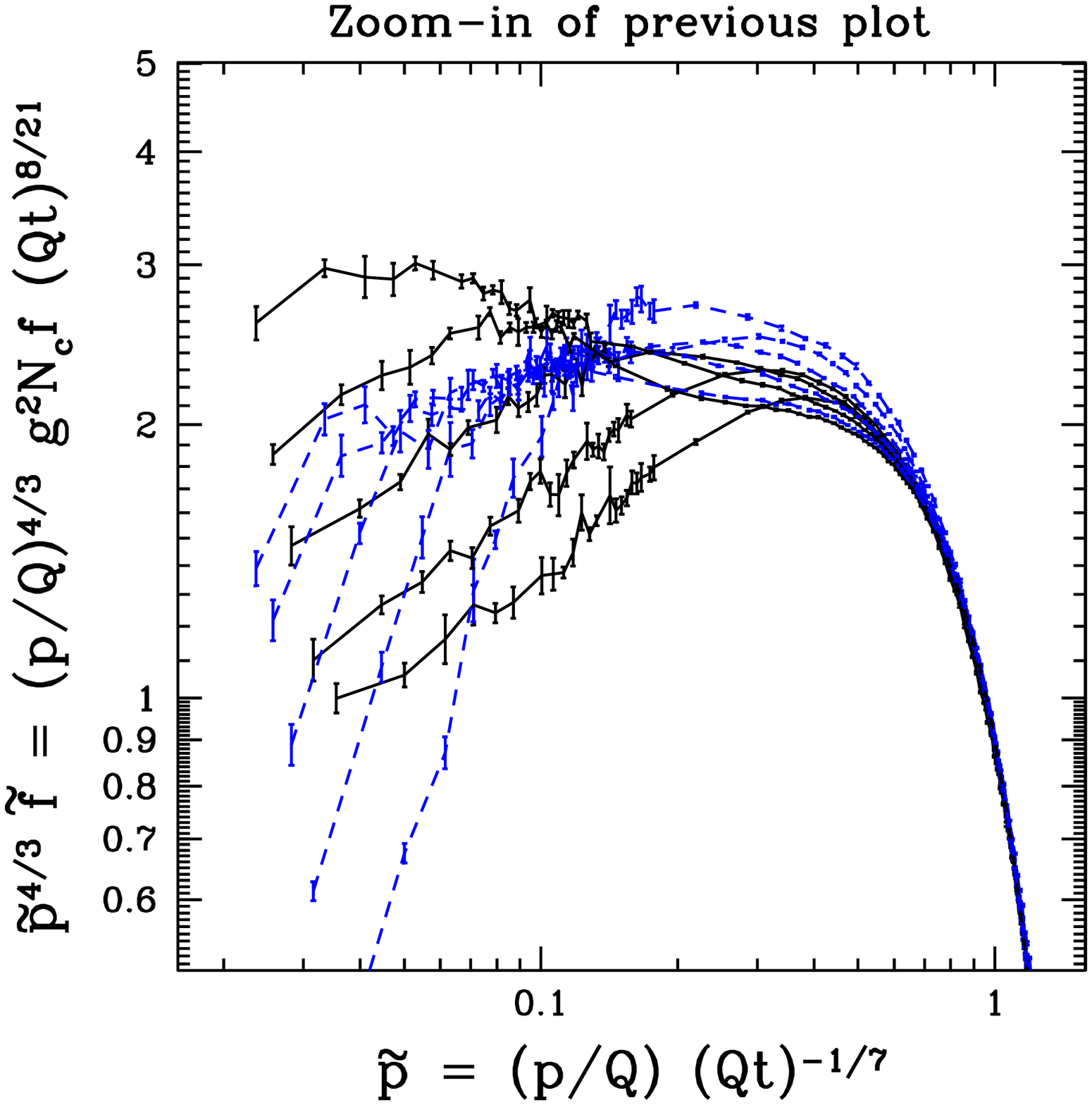} \vspace{3ex} \\
\putbox{0.45}{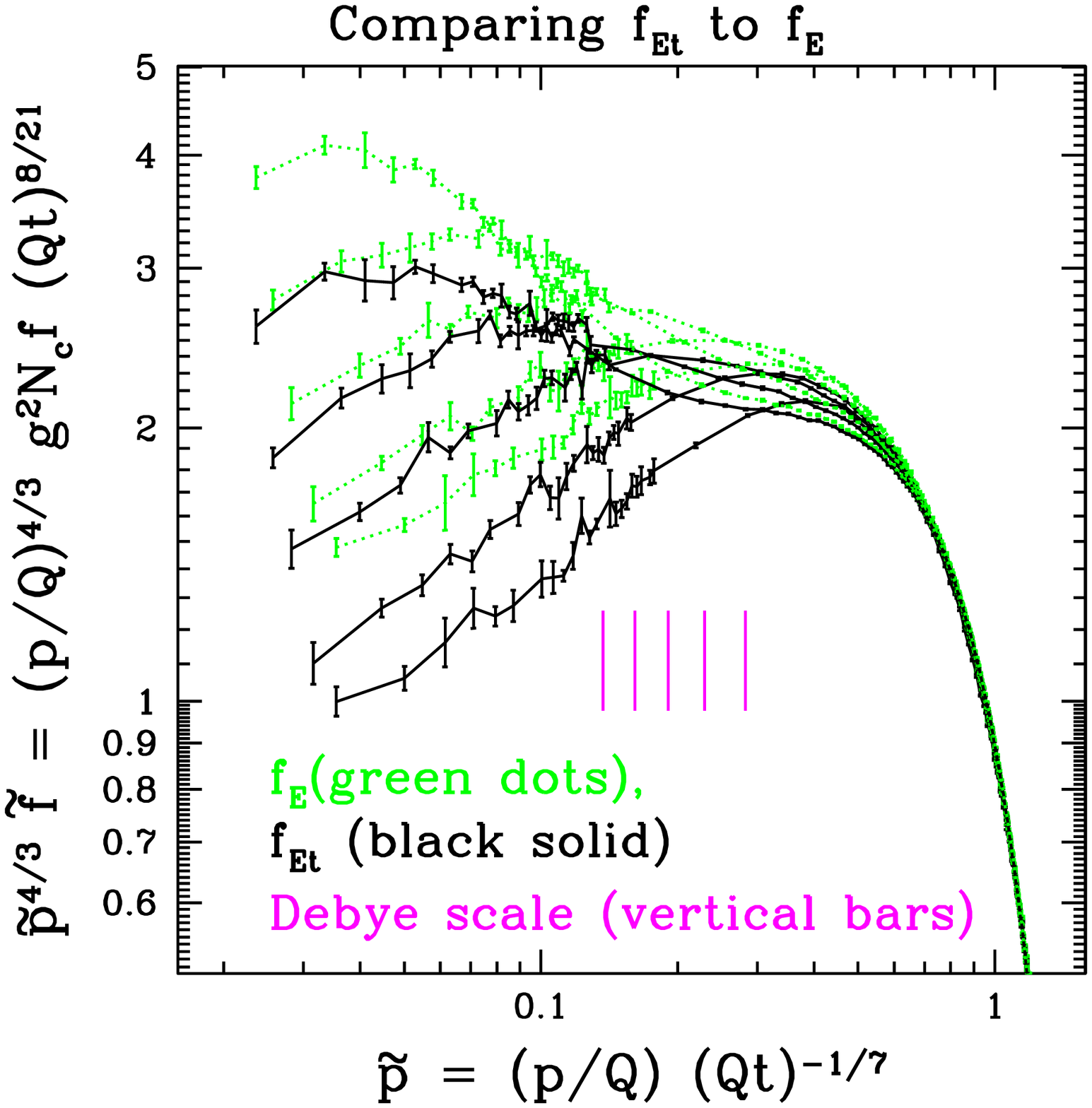} \hfill
\putbox{0.45}{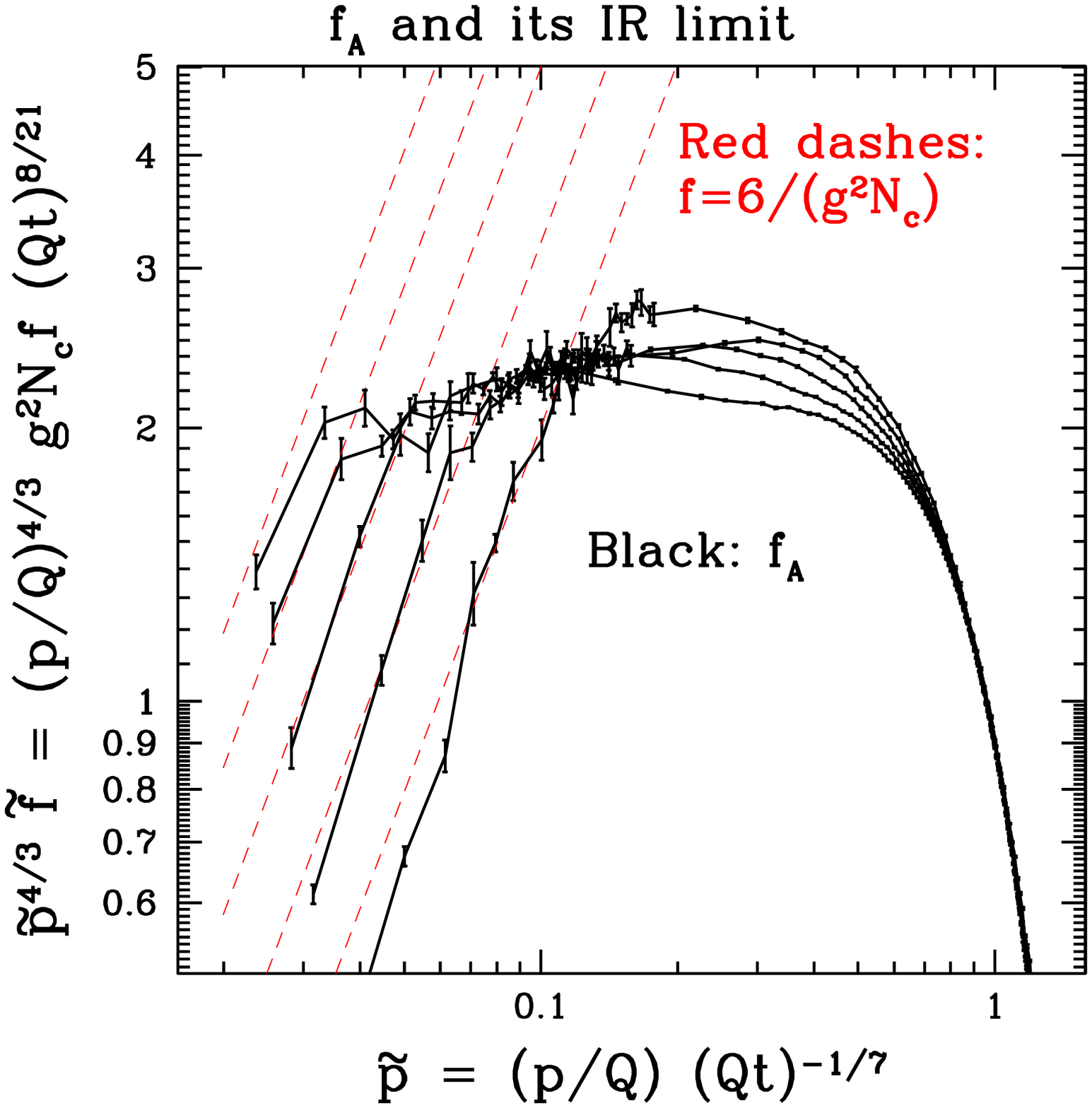}
\caption{
  Occupancies at intermediate times, explicitly scaling out the dominant
  time dependence and the $\ptw^{\frac{-4}{3}}$ momentum dependence,
  based on 7 independent evolutions on $256^3$ lattices.
  Top left:  electric $\fet$ (solid black) versus magnetic $f_A$ (blue
  dashed) occupancies; the earliest time is the bottom curve at
  $\ptw=0.03$ but the top $f_A$ curve at $\ptw=0.4$.
  Top right:  zoom-in of the top-left figure.
  Bottom left: comparison of $\fet$ and $f_E$ (electric occupancies
  contaminated with longitudinal modes, shown in dotted green).
  The (magenta) vertical bars are estimates of the Debye
  mass, with the earliest time at the right.
  Bottom right:  magnetic occupancies, compared to the value where
  $f = 6/g^2 \nc$, the maximum value observed in equilibrium; the lowest
  dashed curve is for the earliest time, the highest for the latest time.}
  \label{fig_midtime}
}

The results are displayed in
Figure \ref{fig_midtime}.  The figure first compares $\fet$ (black solid
lines) to $f_A$ (blue dashed lines), in the top two figures.
To test for a $\ptw^{\frac{-4}{3}}$ infrared scaling behavior,
we have multiplied the occupancies by $\ptw^{\frac 43}$ and plotted
$\ptw^{\frac 43} \ftw(\ptw)$ in the figure.

The lower right plot focuses on $f_A$.  It shows that $f_A$ scales as
$f_A \propto \ptw^{\frac{-4}{3}}$ from  about
$\ptw = 0.4$ out to the infrared scale where the occupancy hits the
value $f = 6/g^2 \nc$, indicated by the red dashed lines.  Then it
saturates at this value.  The value $6/g^2 \nc$ is the same as the
maximum value we found in the equilibrium study, so this appears to
indicate that the occupancy has become nonperturbatively large and
encountered magnetic screening.  We can estimate the total particle
number in nonperturbative infrared magnetic fields by integrating
$f_A(k) d^3 k$ starting where $f_A(k) > 4/g^2 \nc$.  We find
that the fraction of particle number in the condensate,
\be
\frac{n_{\rm condensate}}{n_{\rm total}}
\equiv
\frac{\int_{f_A>4/(g^2 \nc)}  k^2 f_A dk}{\int k^2 f_A dk}\,
\ee
is $4.3\%$, $1.9\%$, $1.1\%$, $0.75\%$, and $0.54\%$
for $Qt = 32$, 72, 156, 299, and 544 respectively.

Now consider the electric occupancies.  The lower left figure compares
$\fet$ with $f_E$.  It also presents an estimate of the Debye screening
scale, obtained by evaluating
\be
\label{md_est}
\mD^2 = 4 g^2 \nc \int \frac{d^3 k}{(2\pi)^3} \frac{f(k)}{k} \,,
\ee
the quantity which would perturbatively give $\mD^2$.  This
determination of $\mD^2$ is indicated for each evolution by a (magenta)
vertical bar, with the rightmost (large-$k$) bar representing the
earliest time and the leftmost (small-$k$) bar the latest time.  As we
have seen, $\mD$ is the scale where we would expect $f_E$ and $\fet$
to start differing from each other significantly, and this is indeed the
case.  Since $f_E$ is established assuming there are only two occupied
polarizations, while below $\mD^2$ there are really three, it is
$\fet$ which should be used as an estimate of occupancy.  The fact that
$f_E$ rises above $\fet$ could lead to an incorrect inference that at
least the electric occupancies rise more steeply than $p^{\frac{-4}{3}}$
in the infrared.  In fact $\fet$ rises in the IR for
$Qt=544$, but not as strongly as we would find using $f_E$.

The use of $\fet$ to estimate occupancies is also
not really correct below the scale $\wpl = \mD/\sqrt{3}$,
because \Eq{f_and_E} and \Eq{E_and_f} assume free dispersion of
plasmons, while the plasmon dispersion is strongly modified in this
region; the factor $1/|\p|$ in \Eq{E_and_f} should presumably be
replaced by $1/\sqrt{p^2 + \wpl^2}$.  (This is also not quite
right because part of the electric field strength resides in the Landau
cut.  We have no good way of separating the pole and cut contributions,
but for $k < \wpl$ one expects the pole (plasmon) contribution to
dominate, so we will ignore this complication.)  If we bravely redefine
$\fet$ as
\be
\fet[\mbox{plasmon}] =  \frac{{\cal P}^{ij}_{\rm T} \delta_{ab}}
            {2(\nc^2{-}1) \sqrt{p^2+\wpl^2}}
 \int d^3 x \: e^{i \p \cdot \x}
            \langle E_i^a(x) E_j^b(0) \rangle_{\rm coul}
\ee
and re-plot just $\fet[\mbox{plasmon}]$ against $\ptw$,
we get the occupancies shown in Figure \ref{fig_wpl}.  (The figure adds
two lines at later times $Qt=976,1680$ obtained on a finer lattice with
$Qa=.298$.)

\FIGURE{
\putbox{0.75}{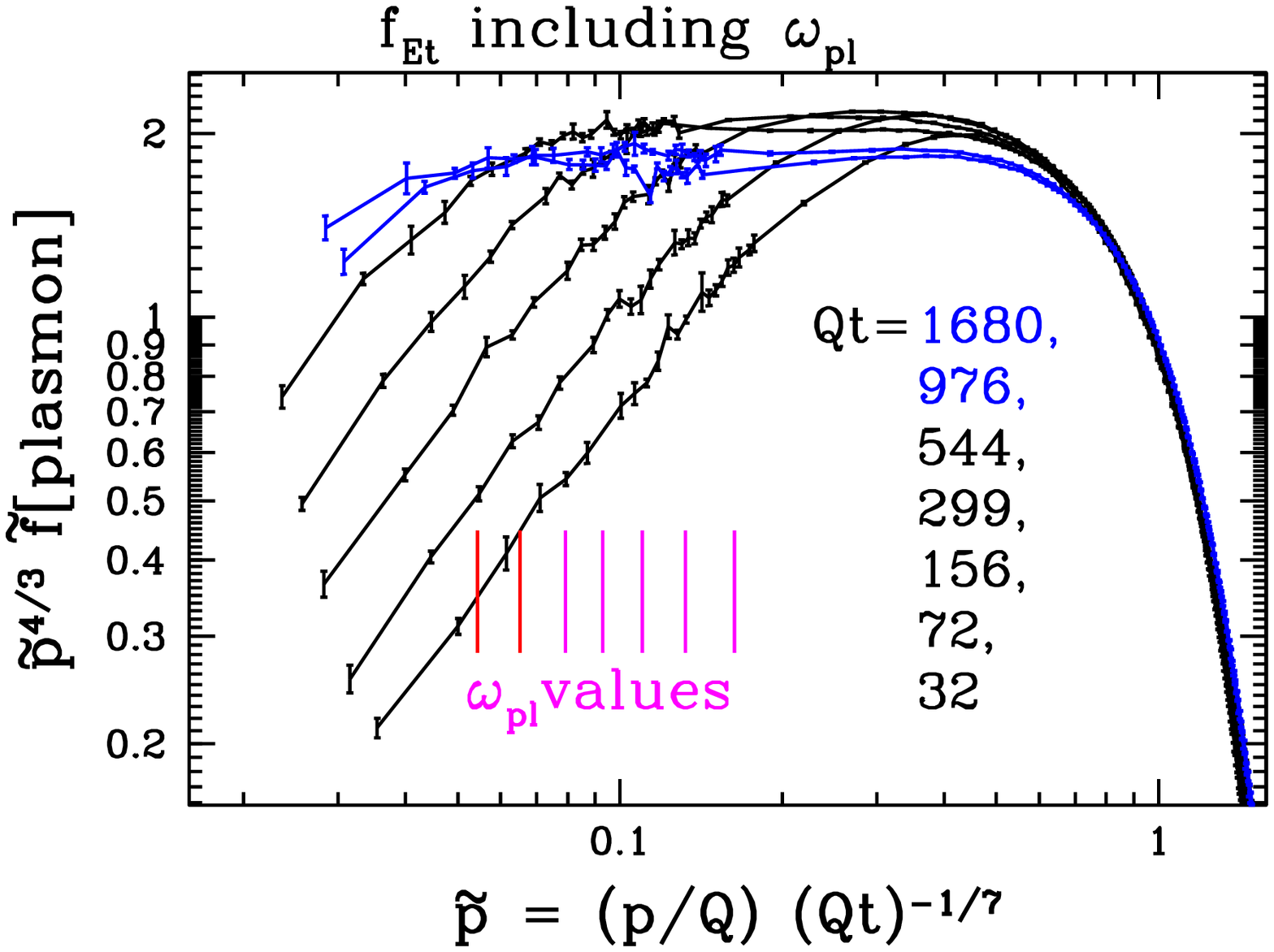}
\caption{
  Occupancies $\fet$ assuming dispersion corrected by the plasma
  frequency.  Data is the same as in Figure \ref{fig_midtime},
  but with two later times (at $Qa=.298$) added.}
  \label{fig_wpl}
}

What Figure \ref{fig_wpl} shows is that the apparent rise in $\fet$,
found in Figure \ref{fig_midtime}, arose because the occupancies were
being mis-evaluated, not taking into account the dispersion of the
plasmons which $\fet$ is describing.  When one takes this into account,
one finds for $Qt>200$ that there is a clean scaling window with
$\ptw^{\frac{-4}{3}}$ scaling behavior, and then a falloff at 
momenta below the plasma frequency.

In summary, as $Qt$ increases, a scaling window opens between the scale
$\ptyp$ and an infrared scale where occupancies become nonperturbatively
large and the physics of plasmons is important.  This scaling window
shows $\ftw \propto \ptw^{\frac{-4}{3}}$.  After correctly identifying
the relation between electric field correlators and occupancy, we find
no evidence for $\ptw^{\frac{-3}{2}}$ scaling at any time.

\section{Are there condensates?}

Recently Blaizot {\it et al} have argued that the evolution of classical
Yang-Mills theory, of precisely the sort considered here, may lead to
the formation of condensates of gluons in the deep infrared
\cite{BGLMV}.  If by condensate one means that the occupancy reaches
the scale $1/g^2$ in the infrared, then we definitely see a
condensate.%
\footnote{Using this definition, there is also a condensate in
  weakly coupled Yang-Mills theory in equilibrium.}
As discussed above, we find that
\be
f(p,t) \sim \frac{1}{g^2 \nc} (Qt)^{\frac{-4}{7}}
          \left(\frac{\ptw}{p} \right)^{\frac 43}
\qquad
\mbox{where}\;\;
\ptw = Q (Qt)^{\frac 17} \,,
\ee
which is easily solved for the scale where $f(p) \sim 1/g^2 \nc$:
\be
f(p,t)\sim \frac{1}{g^2 \nc} \quad \mbox{for} \quad
p \sim Q (Qt)^{\frac{-2}{7}} \,.
\ee
The occupancy integrated at and below this scale is of order
$p^3 f \sim \frac{Q^3}{g^2 \nc} (Qt)^{\frac{-6}{7}}$.  Therefore the
particle number stored in the condensate decays with time as the $-6/7$
power of time.  Relative to the total particle number, which scales as
$(Qt)^{\frac{-1}{7}}$, the condensate makes up a fraction of particle
number which scales as $(Qt)^{\frac{-5}{7}}$.  Using the estimates for
magnetic particle number with $f_A>4/g^2 \nc$ in the last section, we
find the data fit this trend well, with a prefactor of about 0.45 --
that is, if we define as ``condensate'' any modes with occupancy
$f_A > 4/(g^2 \nc)$, we find the condensate makes up
$0.45 (Qt)^{\frac{-5}{7}}$ of the total particle number.

We do not
see evidence for a time-independent or long-lived transient population
of $1/g^2$ occupancy modes in excess of the time-scaling estimate above.
We also do not see evidence for a more robustly-defined condensate,
in which the occupancy in a very narrow momentum range exceeds
$1/g^2$.  We believe that such an $f\gg 1/g^2$ condensate is physically
possible for plasmons, but happens not to occur; and that it is very
difficult for such a condensate even to occur in magnetic fields.  We
will present both of these arguments in more detail in this section.

\subsection{Electric condensates:  decay rate of plasmons}

One possibility is the development of a condensate of plasmons, which
are the low momentum extension of the conventional quasiparticles.  Is
there any evidence for such a condensate in the preceding results on
occupancies?  To answer this question, we think it is useful to see what
a condensate of plasmons would look like and how it would evolve.  We
can do so by artificially introducing a condensate of plasmons into a
classical simulation during the cascade towards the ultraviolet,
examining it with the tools of the previous section.

With this in mind, we have performed three simulations with $Qa=0.42$ on
$256^3$ lattices in which we stop the evolution at some point in time,
fix to Coulomb gauge, add a perfectly spatially uniform electric field,
re-enforce Gauss' Law%
\footnote{%
    Gauss' Law is that $D_i E_i = 0$.  A uniform electric field in
    Coulomb gauge satisfies $\partial_i E_i = 0$, which is not the same;
    therefore the configuration where a uniform $E$ field has been added
    will not satisfy Gauss' Law.  We enforce Gauss' Law by projecting to
    the Gauss constraint using the algorithm of Ref.~\cite{chempot}.},
and then follow the evolution.  We performed three runs, introducing the
electric field at times $Qt = 104$, $Qt = 230$, and $Qt = 440$.

\FIGURE{
\putbox{0.7}{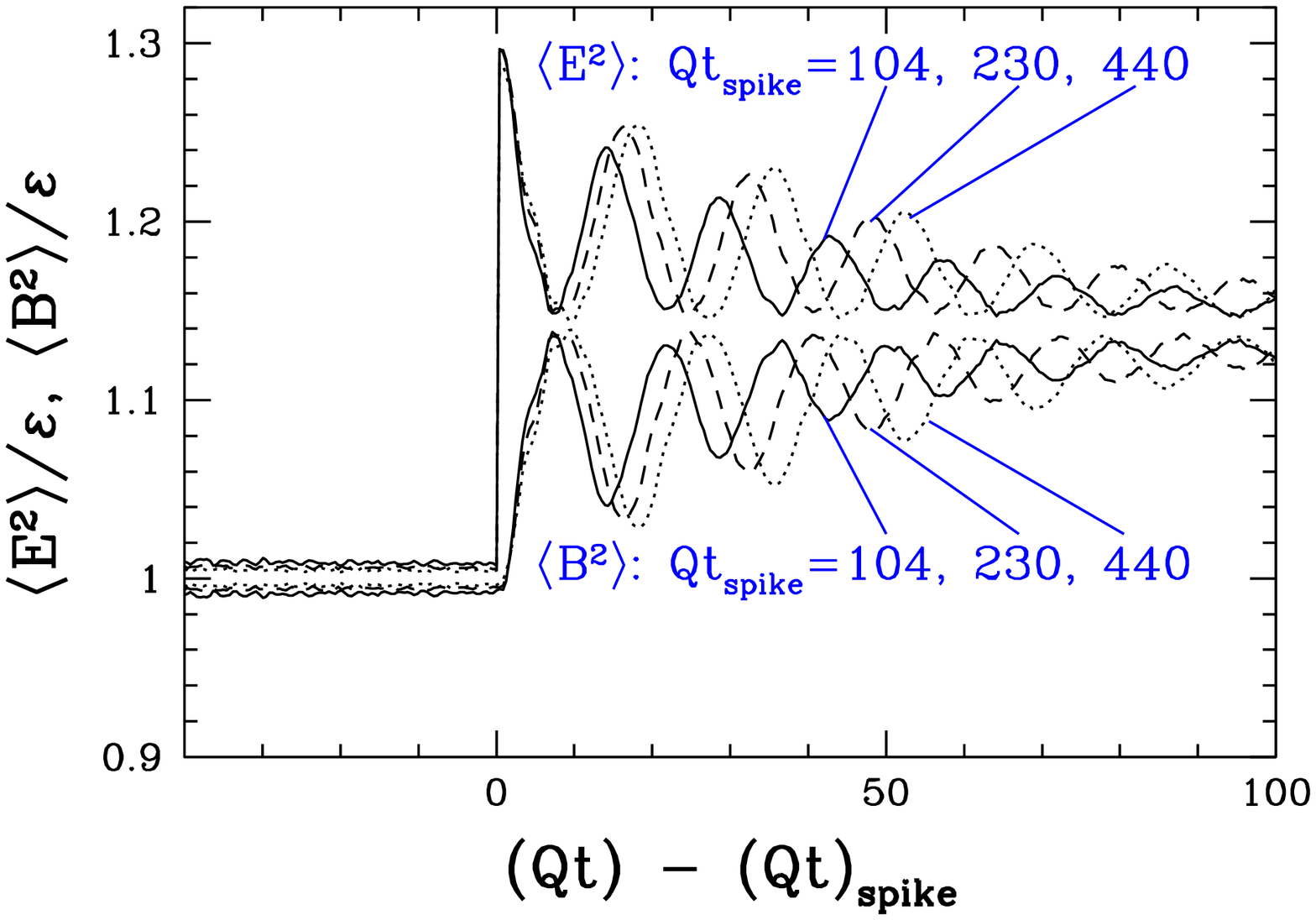}
\caption{
  Electric and magnetic energy as a function of time, when a uniform
  electric field is artificially introduced.  Vertical axis is
  $\langle E^2\rangle$ and $\langle B^2\rangle$ relative to their
  initial average.  At time $(Qt)_{\rm jump}$, a uniform electric field
  is added, leading to (plasmon) oscillations between electric and
  magnetic energy.}
  \label{fig_E1}
}

Figure \ref{fig_E1} plots the electric and magnetic energy as a function
of time after the introduction of the IR electric field.
Unsurprisingly, the electric field energy jumps up at the moment when we
add the coherent $E$-field.  The electric field energy then oscillates
between an almost equal division with magnetic energy and an excess of
electric energy.  These are plasma oscillations.  In a plasma
oscillation, energy goes back and forth between the electric field and a
coherent motion of the charged quasiparticles.  Since quasiparticles
have almost equal electric and magnetic energy, when the plasmon is stored in
quasiparticles, $\avEsq \simeq \avBsq$.  The time between maxima (or
between minima) of the electric field energy is $\pi/\wpl$.
We can estimate $\wpl$ by using $\wpl^2 = \mD^2/3$
and getting $\mD^2$ from \Eq{md_est}, or we can read $\wpl$
directly off the plot; the two agree at the $10\%$ level.  In particular
it is clear that $\wpl$ is smaller at later times.

The amplitude of the oscillations in $E$-field energy decay with time.
This indicates that the plasmons are either scattered to other momenta,
or absorbed by number-changing processes.%
\footnote{%
  If the plasmons were scattered to momenta differing by $\lsim \wpl$
  from the original value, they would lose phase coherence with the
  $k=0$ plasmon, but would continue to have a time-averaged
  $\avEsq > \avBsq$.  This would prevent the
  maxima in $\avBsq$ from almost reaching the
  $\avEsq$ minima.  But we observe that they continue to
  almost meet; so we believe the plasmon damping is dominated by
  absorption and scattering to momenta $k > \wpl$.}
The occupancy falls by about a factor of 2 in one plasma oscillation
($t=2\pi/\wpl$ or two $\avEsq$ peaks), with the decay occurring faster
for the system where we introduced the $E$-condensate at an earlier
$Qt$.  The time scale for plasmon decay is
short compared to the system age for all of the cases we considered.
Therefore we conclude that a condensate of plasmons would be
short-lived; at any time, any plasmons must be of recent origin rather
than quanta left over from the initial conditions.

\FIGURE{
\putbox{0.46}{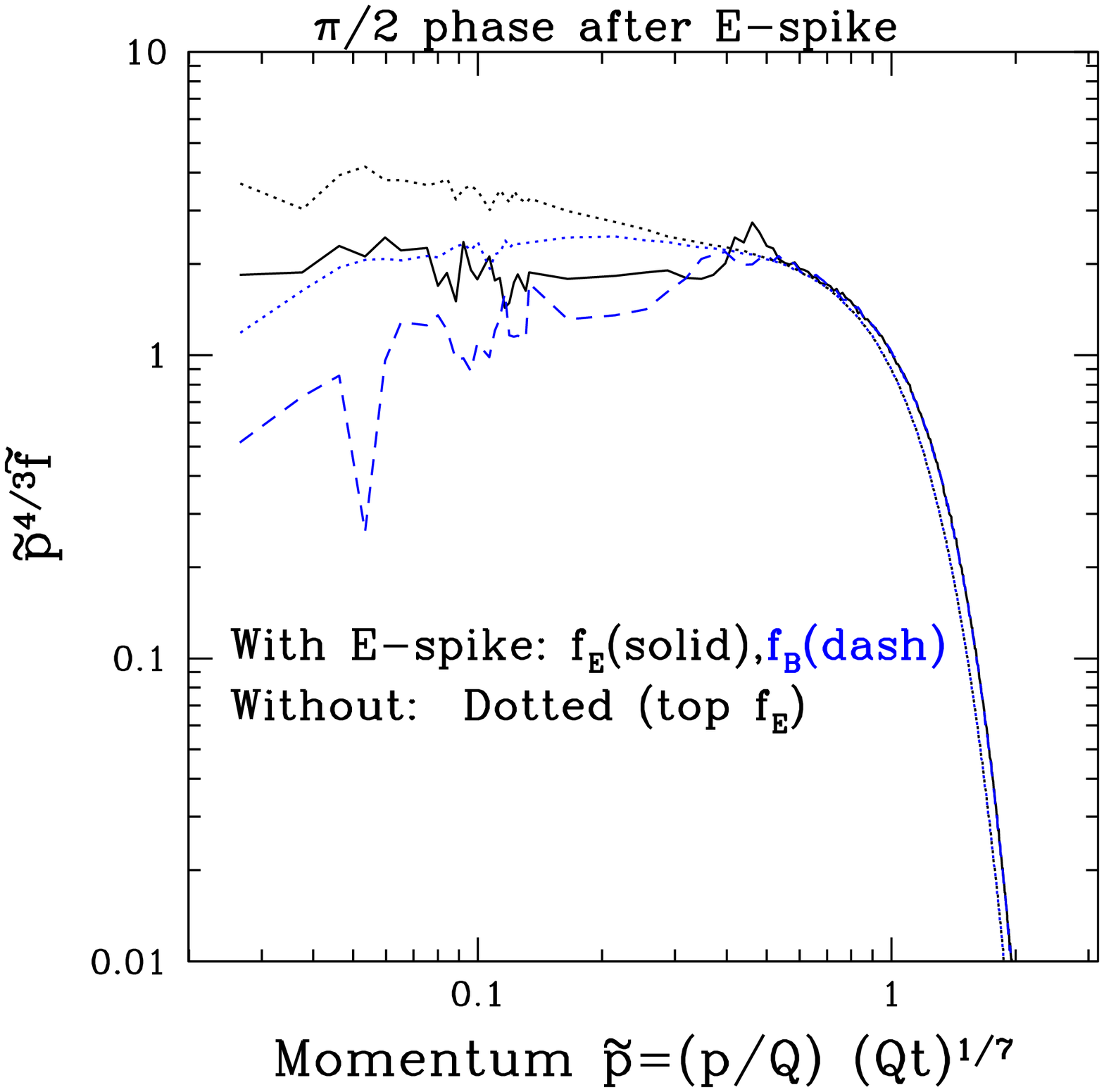}  \hfill
\putbox{0.46}{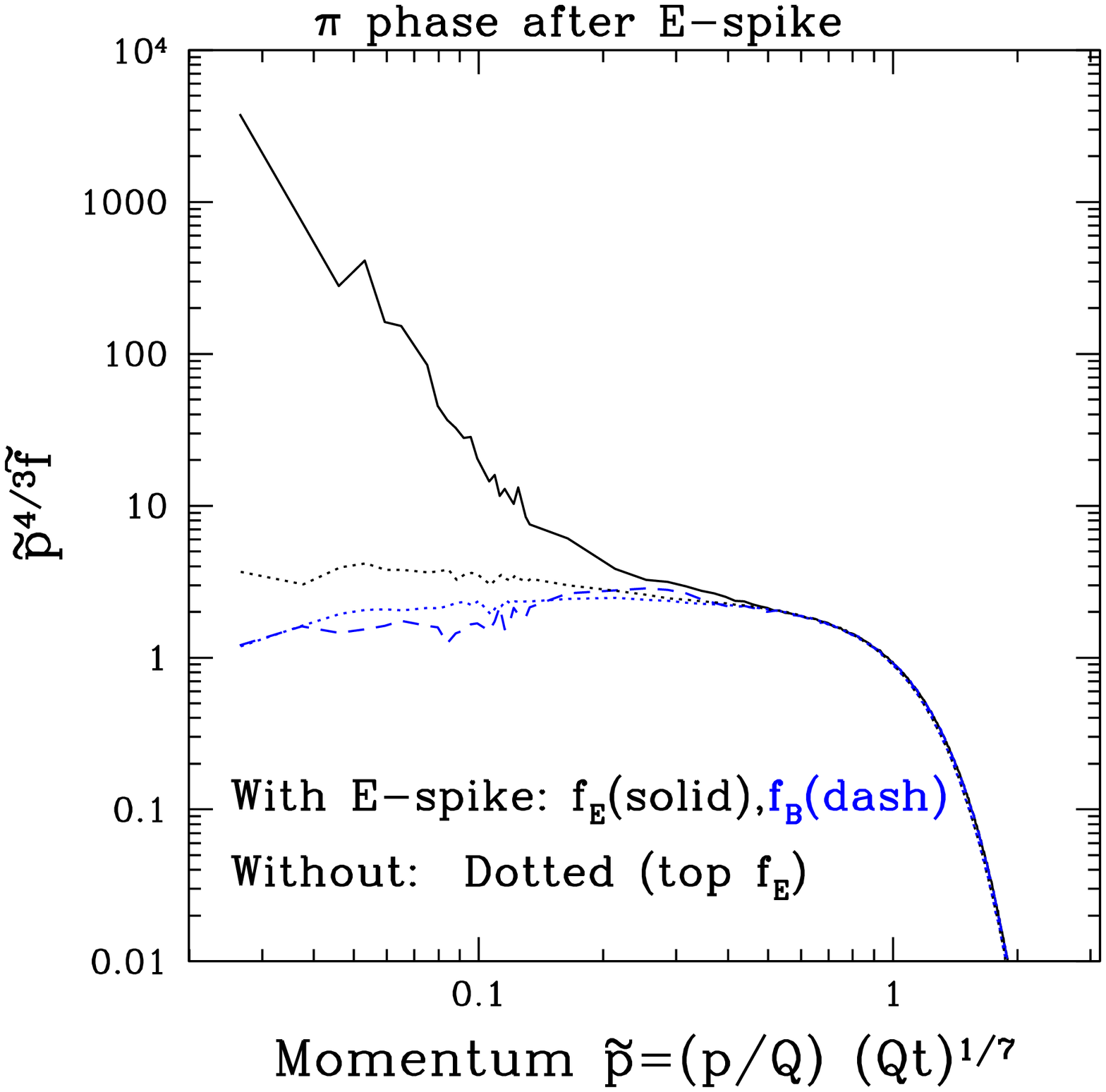}
\caption{
  Left:  comparison of electric occupancy $f_E$ (black) and magnetic
  occupancy $f_A$ (blue) a quarter-oscillation after an $E$-field spike
  is added (first moment when $\avEsq \simeq \avBsq$).  Dotted lines are
  occupancies at the same time, for a run without the $E$-field spike,
  for comparison.  Right:  the same but a half-oscillation after the
  spike, when $\avEsq$ has its next peak.}
  \label{fig_plasmon}
}

We have also investigated the occupancies $f_E$ and $f_A$ when this
large plasmon condensate is present.  Figure \ref{fig_plasmon} shows the
occupancies for our $(Qt)_{\rm spike}=230$ simulation at two points during
the plasmon oscillations:  when the
$E$-field energy has its first minimum and all energy is in
quasiparticles (left), and when the $E$-field has its next maximum and
the energy is in an $E$-field condensate (right).  The dotted lines in
the figure are the occupancies at the same time but in a simulation
where the $E$-field spike is not added, for comparison.
The figure shows that, when $\avEsq \simeq \avBsq$, the plasmon energy
is stored in a redistribution of the quasiparticles.  When $\avEsq$ has
its peak, there is a condensate of soft $E$-fields, with the occupancy
reaching 3 orders of magnitude higher than without the plasmon
condensate.  This is a real condensate!  By comparison, we can conclude
that there is not a condensate of plasmons in the normal simulation.

What this investigation shows is that a condensate of plasmons is
possible in principle.  In particular, an intense and uniform $E$-field
does not create an exponential instability and break up in a time scale
$\tau \sim 1/\sqrt{gE}$ \cite{eatmywords}.  Instead, such a condensate
oscillates between manifesting as an $E$-field and as a coherent motion
of partons.  (Note that the right graph in Figure \ref{fig_plasmon} is
at a time {\sl after} the left graph in the figure, and shows that a
large $E$-field occupancy regenerates from the coherent motion of
quasiparticles.)  

The evidence for
such a condensate in an ordinary simulation would come in two parts; an
excess of electric over magnetic energy, and a spike in the small-$k$
occupancy $f_E$.  A slight excess of $\avEsq$ over $\avBsq$ is observed,
but it is already below $2\%$ for $Qt=50$ and it decays thereafter.
Also, such an excess is expected because of ordinary perturbative
interactions, which should lead to an excess of order
$\mD^2/\ptw^2 \sim (Qt)^{\frac{-4}{7}}$ (the scaling we observe).  We also
definitely do not observe an infrared spike in $f_E$, as shown in Figure
\ref{fig_midtime} and Figure \ref{fig_wpl}.  So while an electric or
plasmon condensate is possible, we do not see one.

\subsection{Magnetic condensates:  Nielsen-Olesen instabilities}

What about condensates in the magnetic sector?  We will argue here that
such condensates are difficult to define or identify in Coulomb gauge.
But we will also show that genuine condensates, in the sense of magnetic
field occupancies greatly in excess of $1/g^2$ for a narrow range of
$k$-vectors, are highly unstable in a nonabelian theory.

\FIGURE{
\putbox{0.42}{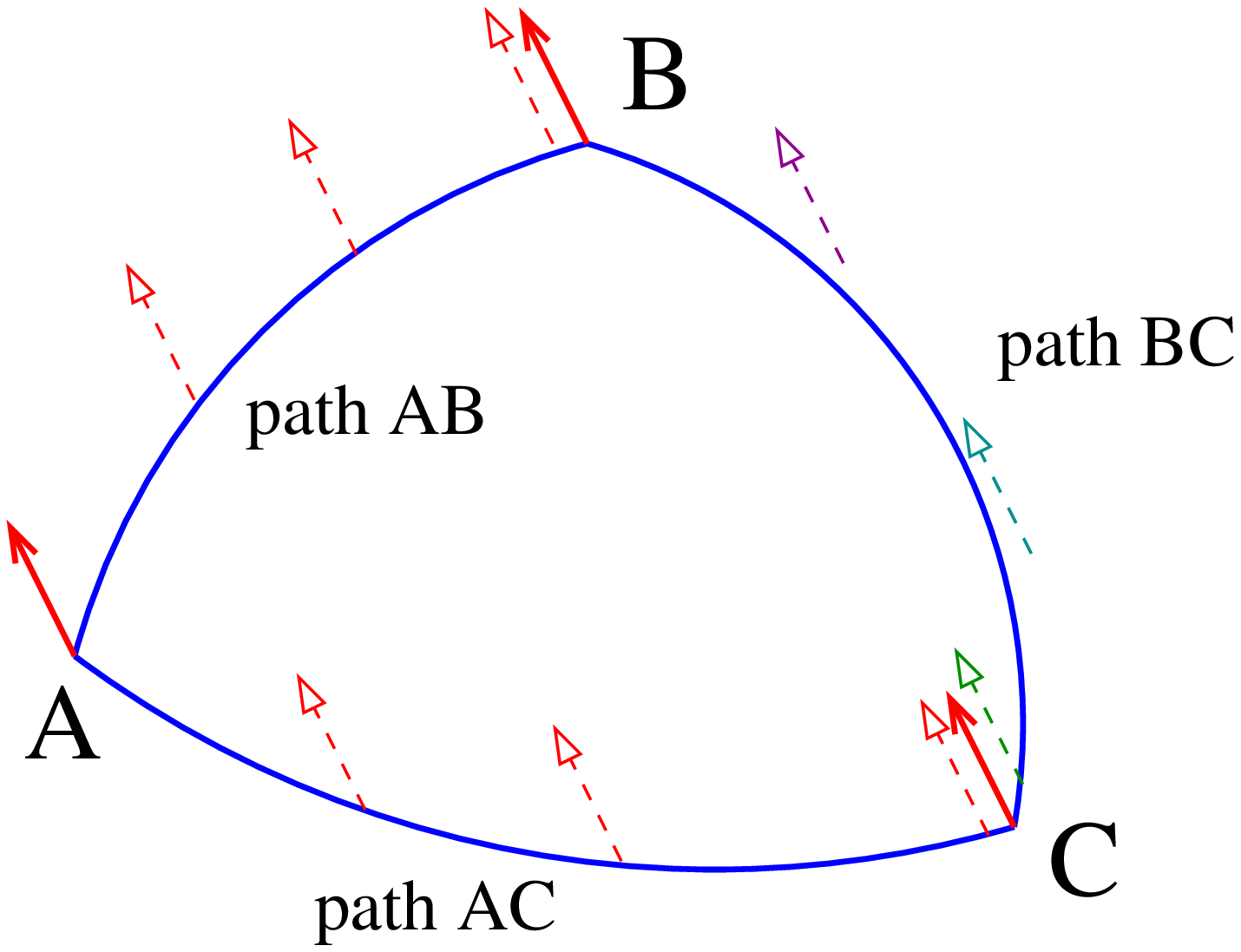}
\caption{
  Illustration of the obstruction to a magnetic condensate:  if parallel
  transport of the field at $A$, along path $AB$, agrees with the field
  at $B$; and transportation along path $AC$ agrees with the field at
  $C$; then since the Wilson loop is very nontrivial, transportation
  from $B$ to $C$ must pick up a large color rotation.}
\label{Wilsonfig}
}

First we explain why it is difficult for magnetic occupancies to be
highly coherent in the sense of occupancies exceeding $f_A\sim 1/g^2$,
or equivalently $\langle A A(k)\rangle \sim k^2/g^2$.
Suppose that there is a gauge field configuration with infrared fields of size
$A(x) \sim 1/(gL)$, with $L$ some length scale.  Consider Figure
\ref{Wilsonfig} and ask if the field can be coherent between points
$A,$ $B$, and $C$, separated by distances of order $L$.  We might say
that the field at $A$ and at $B$ are ``the same'' if the parallel
transport of the field at $A$, along the path $AB$, is the same or
similar to the field at $B$.  Similarly, the field is the same between
$A$ and $C$ if ${\bm A}(A)$ parallel transported along path $AC$ is the
same as ${\bm A}(C)$.  But because the field is large, the Wilson line
along $AB$ and then $BC$ will differ by an order-1 group element from
the Wilson line along $AC$ -- roughly, by $g$ times the flux of
${\bm B}$ field going through the loop, which is order 1.  Therefore,
the parallel transport of the field at $A$, first along path $AB$ and
then along path $BC$, {\sl cannot} agree with the same field parallel
transported just along path $AC$.  So a comparison of the field at $B$
and at $C$, using the path $BC$, will show that they are very
different.  Therefore there is no sense that the
field is really coherent on scales as large as or larger than $L$.

Since there is no gauge invariant sense in which a gauge field of
amplitude $|{\bm A}|\sim 1/gL$ can be coherent on scales $\gsim L$,
it would be surprising -- and probably just a gauge choice
artifact -- for ${\bm A}$ to be coherent on longer scales.
When a field is random over scales longer than $L$, its Fourier
transform has roughly uniform power over momenta less than $k\sim 1/L$.
If the real-space field strength is $A^2 \sim 1/(g^2 L^2)$ and this is
distributed over Fourier modes with $k \sim 1/L$, then the Fourier-space
correlator will be $\langle A^2(k) \rangle \sim k/g^2$, corresponding
to an occupancy $f_A \sim 1/g^2$ for $k \lsim 1/L$ (see
\Eq{f_and_A}).

A loophole in the above reasoning is the case where ${\bm A}$ points in
the same color direction throughout a large region of space -- that is,
if ${\bm A}$ happens to be approximately abelian (or more generally, if
${\bm A}$ commutes with the field strength through any arbitrary Wilson
loop).  This is very
nongeneric, since ${\bm A}$ is specified by $3(\nc^2{-}1)$ numbers (each
color can point in a different space direction), but a gauge
transformation contains only $(\nc^2{-}1)$ independent elements.
Therefore $2(\nc^2{-}1)$ elements of ${\bm A}$ must coincidentally
vanish for it to be nearly abelian. Nevertheless it is useful to
consider what happens in this special case, where there certainly can be
coherence on large scales.  So consider the case where only the $\tau^3$
component of ${\bm A}$ is nonzero, and ${\bm B}$ is uniform in space.
It is quite easy to consider this possibility on the lattice, and we
have done so, adding roundoff error sized fluctuations in other field
modes. We observe the exponential growth of electric field fluctuations;
we plot $\avBsq$ and $\avEsq$ as a function of $Qt$ in Figure
\ref{NO_EB}, and $f_A$ and $f_E$ as a function of $p/Q$ at a series of
times in Figure \ref{fig_NO}.  (In both figures we set the zero of time
to the first moment when $\avEsq=\avBsq$.)
Figure \ref{NO_EB} shows that the electric field
energy grows exponentially until it actually exceeds the magnetic
energy; then there are brief, highly damped oscillations, ending in an
almost equal energy partition between $E$ and $B$.  The lefthand plot of
Figure \ref{fig_NO}  shows occupancies during the process of electric
field growth, with successive curves at successive times separated by
about $Qt=0.8$.  The righthand plot shows the process of field
saturation at a series of times around $t=0$.  The figure shows that the
magnetic condensate (the spike at the earliest time) disappears; then
there is some chaotic dynamics with a very short-lived high IR occupancy
of electric fields.  But these fields are mostly gone by $Qt=3.2$, when
the configuration already resembles the occupancies during the cascade.

\FIGURE{
\putbox{0.46}{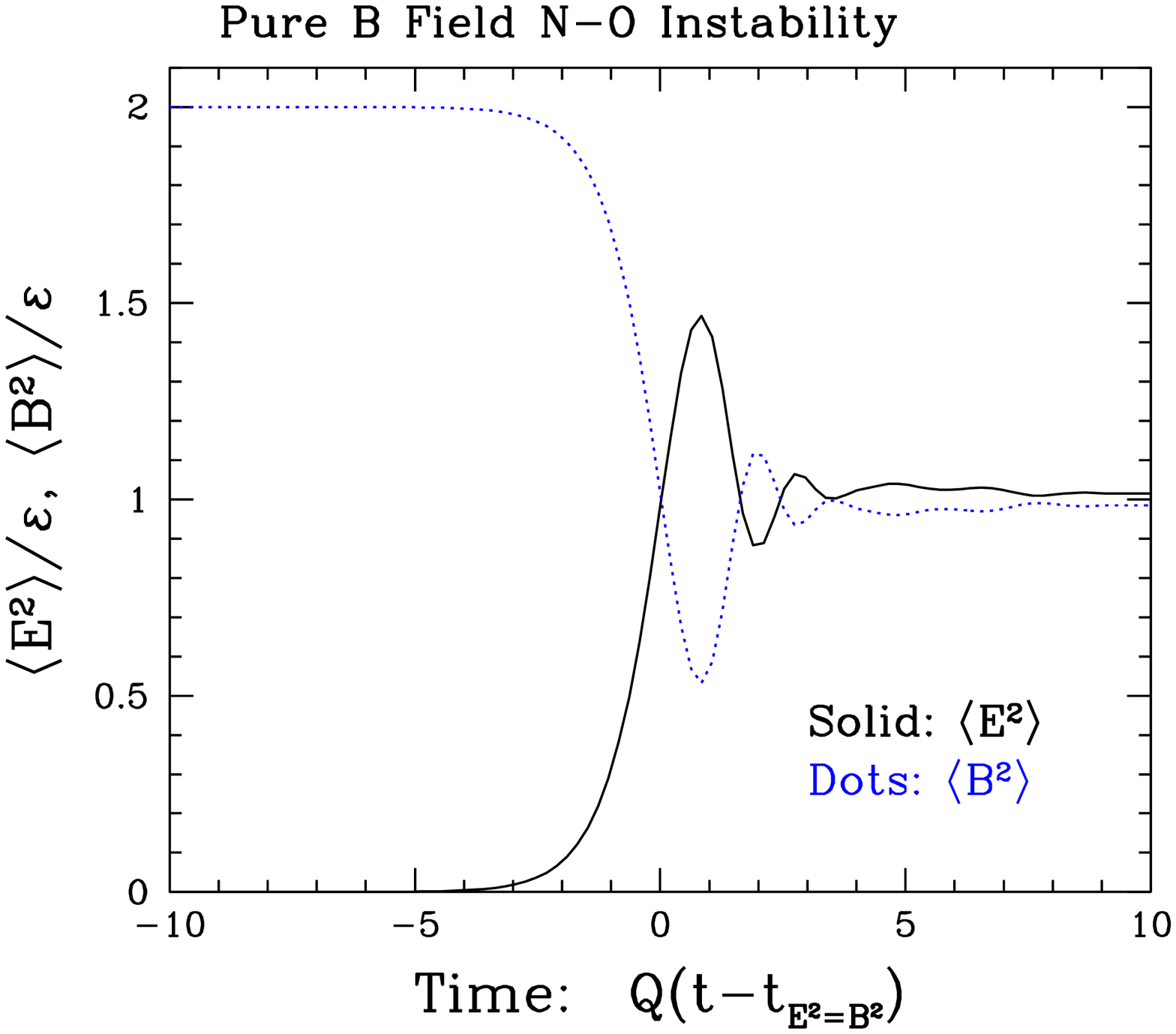} \hfill
\putbox{0.46}{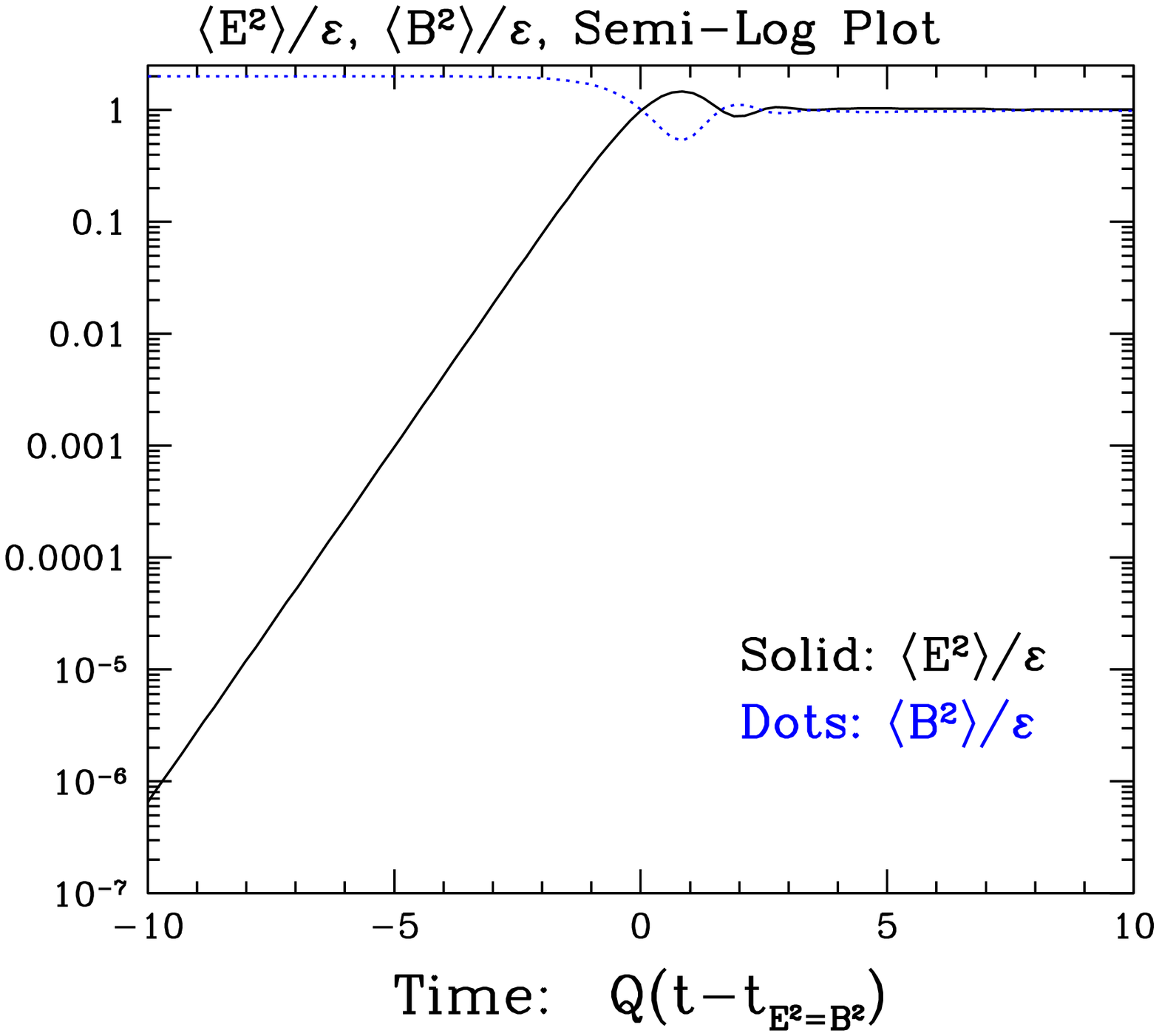}
\caption{
  Electric and magnetic energy fractions $2 \avEsq/(\avEsq+\avBsq)$,
  $2\avBsq/(\avEsq+\avBsq)$ as a function of time during Nielsen-Olesen
  instabilities.  Left:  linear plot.  Right:  log-linear plot.}
  \label{NO_EB}
}

\FIGURE{
\putbox{0.46}{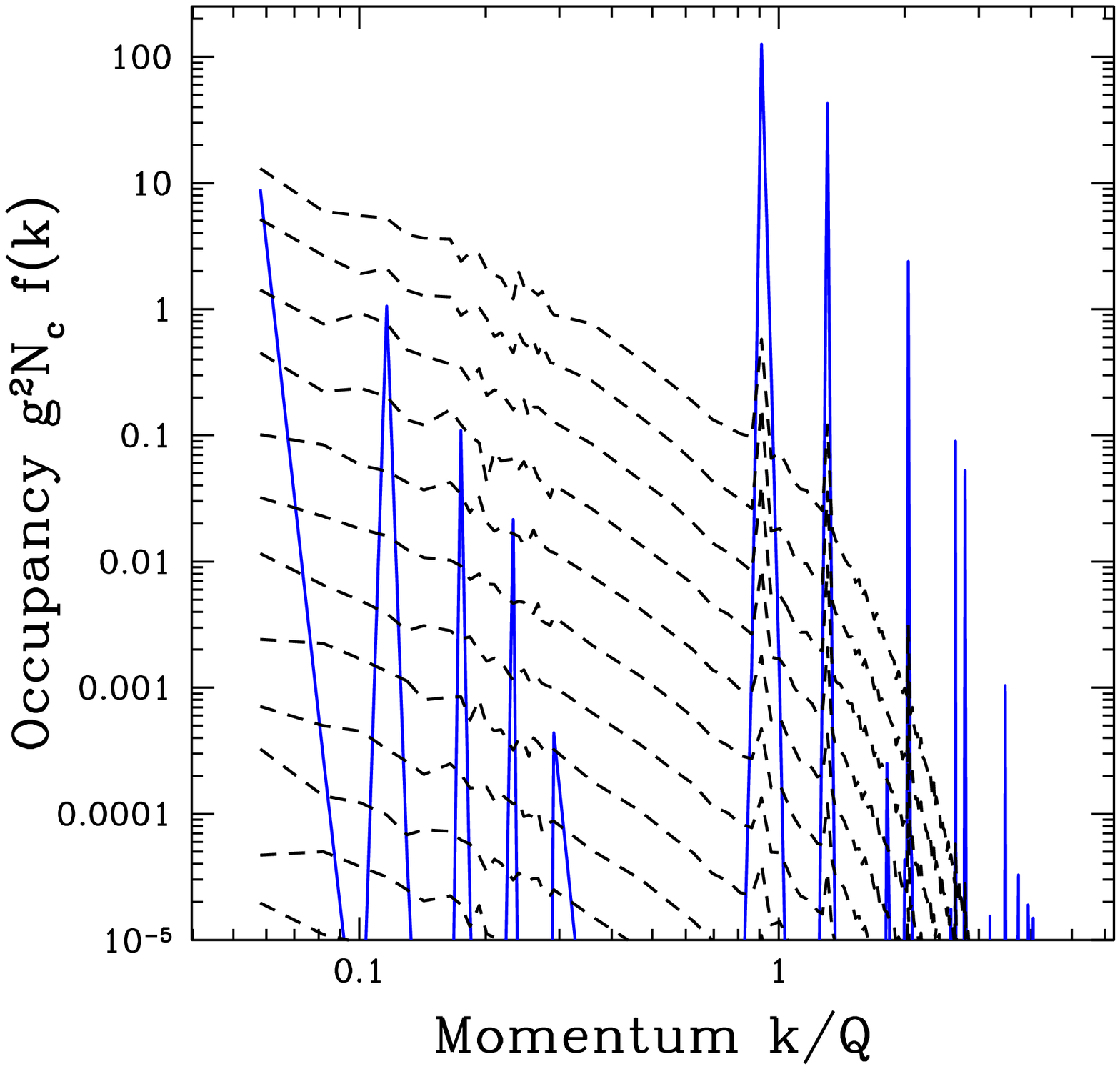}
\hfill
\putbox{0.46}{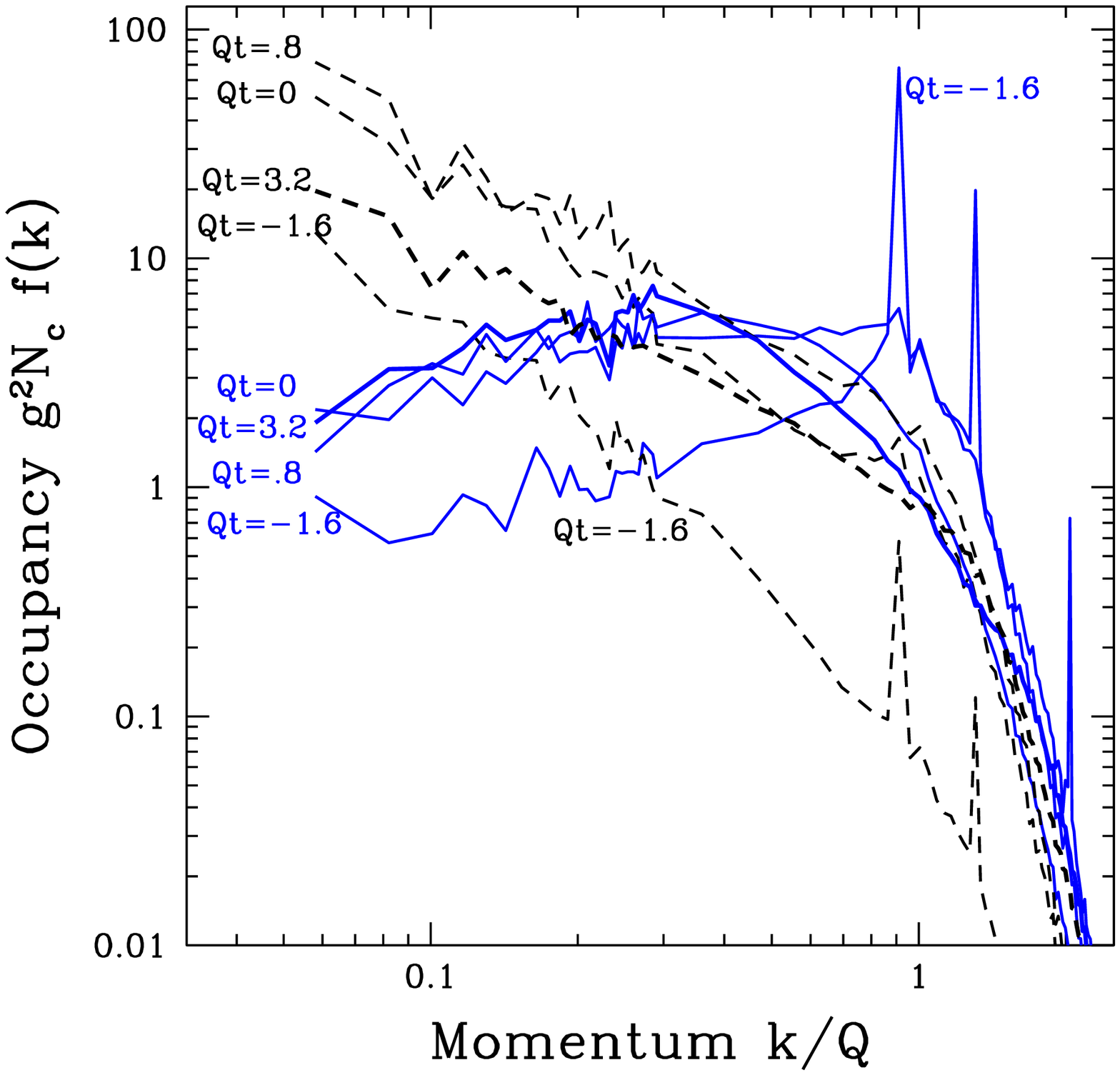}
\caption{
  Growth of Nielsen-Olesen instabilities for a uniform abelian
  $B$-field.  Left:  the electric occupancies (dashed lines) grow in the
  presence of the magnetic condensate (solid, spiky line).  Right:  the
  $B$-field condensate disappears,
  there is briefly a very large $E$-field occupancy in the infrared, and
  then the configuration becomes the first stage of the usual cascade to
  the ultraviolet.}
  \label{fig_NO}
}

The reason for the exponential growth of electric fields in the presence
of a uniform $B$ field is the Nielsen-Olesen instability.  As shown by
Nielsen and Olesen \cite{NO}, in a uniform $B$ field there are modes
whose amplitude will grow at the rate%
\footnote{%
    Here is a lightning quick derivation of this result.  In the
    presence of a magnetic field, charges follow Lamor orbits, causing
    transverse momentum to be quantized,
    $k_\perp^2 = gB(1+2n)$, with $n=0,1,2,\ldots$ the Landau level.  The
    energy of an excitation is
    $\epsilon = \sqrt{k_\perp^2 - 2g S\cdot B + k_\parallel^2}$, with
    $-2g S\cdot B$ spin-magnetic interaction term.
    For a spin-1 particle, $-2gS \cdot B = \pm 2gB$ depending on
    polarization.  The minimum of $\epsilon$ is obtained
    when $n=0$, $k_\parallel=0$ and $-2gS\cdot B = -2gB$.  In this case
    the energy is $\epsilon = \sqrt{-gB}$, corresponding to exponential
    growth with exponent $\exp(t \sqrt{gB})$.}
\be
|{\bm E}| \propto \exp(+\gamma t) \,, \qquad
\gamma = \sqrt{g B} \,.
\ee
The occupancy and energy go as $E^2$ and grow at twice this rate.  This
is precisely the growth rate we observe.  The arguments of Nielsen and
Olesen apply whenever the magnetic field is coherent over a distance
scale large compared to $R_{\rm Lamor} = 1/\sqrt{gB}$.  This essentially
forbids magnetic fields with such coherence as to have $f_A \gg 1/g^2$;
any such magnetic field would be exponentially unstable and rapidly
dissolve.

A curious feature of Figure \ref{fig_NO} is that, even though we started
with a perfectly uniform and abelian gauge field configuration, the
process of finding Coulomb gauge has interpreted it as a nonuniform and
nonabelian configuration; the occupancy $f_A$ in Figure \ref{fig_NO}
has its peak at $k\sim Q$ rather than at $k=0$.
It is still recognizable as a condensate; using our former definition
that ``condensate'' means $f_A > 4/(g^2 \nc)$, we find about $90\%$ of
the particle number is in a condensate.  The fact that the field
is highly coherent is stored in the very peculiar structure of the
inferred occupancies, with a sharp spike reaching far above
$f=6/g^2 \nc$, the limit we have observed in thermal and stochastic
cascading systems.
We believe that a genuine magnetic condensate would
appear as a similar, structured and sharp feature in the magnetic
occupancy -- a feature which does not appear at any point in our
simulations except those which start with a coherent $B$-field.

To summarize, we directly observe the Nielsen-Olesen instability;
whenever the magnetic field $B$ is coherent on a scale larger than the
Lamor radius of charges in the field, $R_{\rm Lamor}  = 1/\sqrt{gB}$,
there are exponentially unstable modes, growing on a time scale
$\gamma = \sqrt{gB}$, which consume the $B$-field and turn into a
stochastic collection of excitations.  This physics should prevent
fields with $k<\sqrt{gB}$, corresponding to occupancies larger than
roughly $f_A \gsim 2\pi/g^2$.

\section{Discussion}

Classical Yang-Mills theory, initially occupied in the infrared, sees a
cascade of energy towards the ultraviolet, due to the infinite phase
space available there.  Simple arguments, which suggest
$\ptyp \sim Q (Qt)^{\frac 17}$ and
$g^2 \nc\, f \sim (Qt)^{\frac {-4}{7}}$, are borne out, and in fact,
scaling out this dominant behavior, we find that the occupancy $f(p,t)$
rapidly approaches a scaling solution.  

The infrared dynamics at
intermediate times ($Qt\sim 100$) are complicated because the screening
and nonperturbative scales are not well separated from the scale
$\ptyp$.  At the scale $\mD \sim \ptyp (Qt)^{\frac{-2}{7}}$ the electric
field takes on significant longitudinal occupancy, and one should use
only its transverse components to estimate the quasiparticle occupancy,
to avoid miscounting.  Below around $\wpl = \mD/\sqrt{3}$ the
dispersion of plasmons should also be considered.  Taking these effects
into account, we find no evidence for $p^{\frac{-3}{2}}$ scaling; when
any scaling window exists in the infrared, it always shows exponent
$p^{\frac{-4}{3}}$.  The occupancy determined from $A$-field correlators
agrees with that from $E$-field correlators until around the scale
$\mD$, where they start to diverge.  The magnetic occupancy $f_A$
saturates at about $f_A = 6/(g^2 \nc)$, and the total particle number in
such nonperturbatively large fields (magnetic condensate) scales with
time as
$\frac{n_{\rm condensate}}{n_{\rm tot}} \sim 0.45 (Qt)^{\frac{-5}{7}}$.

It is possible for electric (plasmon) condensates to exist; indeed, we
can artificially introduce them into a lattice evolution to investigate
their behavior.  The plasmons can carry occupancy large compared to
$1/g^2$, but in practice they do not.  And the damping rate for plasmons
is fast compared to the system's age.

Genuine magnetic condensates are unstable due to the Nielsen-Olesen
instability; we verify the growth rate of fluctuations in the presence
of a perfectly coherent $B$-field, $\gamma = \sqrt{gB}$.  The
high-occupancy infrared magnetic field observed during the scaling
cascade to the ultraviolet better resembles the nonperturbative
incoherent $B$-fields in equilibrium weakly-coupled Yang-Mills theory.

It would be interesting to solve the Boltzmann equations for Yang-Mills
theory, presented in Reference \cite{AMY5}, directly, to compare against
the numerical scaling solution found here.

\section*{Acknowledgments}
We would like to thank Kari Rummukainen and Mark York for collaboration
during the early stages which lead to this paper, and
J\"urgen Berges, S\"oren Schlichting,
Francois Gelis, and Raju Venugopalan for conversations.  We also thank
the Department of Energy's Institute for Nuclear Theory at the
University of Washington for providing the engaging
environment in which this work was started.  This work was supported in
part by the Canadian National Sciences and Engineering Research Council
(NSERC) and the Institute of Particle Physics (Canada).  Part of the
computational work was performed at the Finnish IT Center for Science
(CSC), Espoo, Finland.

\appendix

\section{Equilibrium $E$-field correlators}

\label{App1}

Consider classical Yang-Mills theory with some UV regulator (such as the
lattice).  Because the theory is UV regulated it has a well defined
thermal ensemble, described by the path integral
\be
{\cal Z} = \int {\cal D}(A_i,E_i) \;
             \exp\left( -\frac{E_i E_i + B_i B_i}{2T} \right)
             \delta(D_i E_i) \,.
\label{classicalZ}
\ee
The delta function enforcing Gauss' Law is a functional delta, with one
such delta at each point.  We can rewrite it in terms of a Lagrange
multiplier, suggestively named $A_0$ \cite{AmbjornKrasnitz}:
\be
\delta(D_i E_i) = \int {\cal D}A_0 \;
      \exp\left( \frac{i A_0 D_i E_i}{T} \right) \,.
\label{A0}
\ee
If we perform the Gaussian integration over the $E$-field we obtain
\be
\int {\cal D} (A_0,E_i) \;
\exp\left( \frac{-E_i E_i +2 i A_0 D_i E_i}{2T} \right)
= \int {\cal D} A_0 \;
 \exp\left( \frac{-(D_i A_0)^2}{2T} \right) \,,
\label{A0kinetic}
\ee
the standard bare kinetic term for the $A_0$ field.  Under
renormalization this mixes with the identity, generating the Debye mass
\cite{Kajantie}:
\be
(D_i A_0)^2 \; \Rightarrow_{\rm IR} \;
(D_i A_0)^2 + \mD^2 A_0^2 \,.
\label{and_mD}
\ee
But this IR behavior should already hold even if we don't integrate out
the $E$-field.  In the infrared, the behavior of ${\cal Z}$ is
\be
{\cal Z}_{\rm IR} = \int {\cal D}(A_i,E_i,A_0) \;
  \exp \left( - \frac{E_i E_i + B_i B_i -2i A_0 D_i E_i + \mD^2 A_0^2}
       {T} \right) \,.
\label{Z2}
\ee
Now we can integrate out the $A_0$ field,
\be
\label{Z3}
{\cal Z}_{\rm IR} = \int {\cal D}(A_i,E_i) \;
  \exp \left( - \frac{E_i E_i + B_i B_i + \mD^{-2} (D_j E_j)^2 }
       {T} \right) \,,
\ee
neglect the difference between $\partial_j$ and $D_j$, and use this to
evaluate the transverse and longitudinal $E$-field correlators:
\bea
\hat{k}_i \hat{k}_j \langle E_i E_j(k) \rangle &=& \frac{T}{1+k^2/\mD^2}
\,, \\
\half {\cal P}_{ij} \langle E_i E_j(k) \rangle & = & T \,.
\eea
Combining with \Eq{Et_and_f} and \Eq{El_and_f}, \Eq{sumrule} follows.
The sum rule is not exact because we have approximated $D_i E_i$ with
$\partial_i E_i$; but our numerical results in Figure \ref{fig_therm}
show that it is surprisingly accurate.


\begin{thebibliography}{29}
\bibitem{eatmywords}
  Contrary to what one of us (GDM) has claimed in discussions on some
  recent occasions.
\end{thebibliography}
\end{document}